\documentclass[letterpaper,twocolumn,10pt]{article}
\usepackage{zhanggroup}

\usepackage{hyperref}
\hypersetup{
colorlinks,
linkcolor={blue!70!green},
citecolor={green!70!blue},
urlcolor={orange!70!red}
}
\usepackage{xspace}
\usepackage{cite}
\usepackage{url}

\usepackage{amsmath,amssymb,amsfonts}
\usepackage{algorithmic}
\usepackage{graphicx}
\usepackage{textcomp}
\usepackage{xcolor}
\usepackage{tikz}
\usepackage{array}
\usepackage{pifont}
\usepackage[normalem]{ulem}
\usepackage{multirow}
\usepackage{subcaption}
\usepackage{caption}
\usepackage{booktabs}
\usepackage{makecell}
\usepackage{tcolorbox}
\usepackage[absolute]{textpos}

\graphicspath{ {images/} }
\usepackage{enumitem}
\setlist[itemize]{leftmargin=*}

\newcommand{\mypara}[1]{\smallskip\noindent\textbf{#1.}\xspace}

\newcommand{\mybox}[1]{
\begin{tcolorbox}[boxsep=1pt,left=2pt,right=2pt,top=0 pt,bottom=0.5pt, frame empty]
\emph{Takeaways:} #1
\end{tcolorbox}
}
\begin{document}

\begin{textblock}{13}(1.5,1)
\centering
To Appear in the 32nd USENIX Security Symposium, August 2023.
\end{textblock}

\title{\Large \bf A Plot is Worth a Thousand Words: Model Information Stealing Attacks via Scientific Plots}

\date{}

\author{
Boyang Zhang\textsuperscript{1}\ \ \
Xinlei He\textsuperscript{1}\ \ \
Yun Shen\textsuperscript{2}\ \ \
Tianhao Wang\textsuperscript{3}\ \ \
Yang Zhang\textsuperscript{1}\ \ \
\\
\\
\textsuperscript{1}\textit{CISPA Helmholtz Center for Information Security}\ \ \
\textsuperscript{2}\textit{NetApp}\ \ \
\textsuperscript{3}\textit{University of Virginia}
}

\maketitle

\begin{abstract}

Building advanced machine learning (ML) models requires expert knowledge and many trials to discover the best architecture and hyperparameter settings.
Previous work demonstrates that model information can be leveraged to assist other attacks, such as membership inference, generating adversarial examples.
Therefore, such information, e.g., hyperparameters, should be kept confidential.
It is well known that an adversary can leverage a target ML model's output to steal the model's information.
In this paper, we discover a new side channel for model information stealing attacks, i.e., models' scientific plots which are extensively used to demonstrate model performance and are easily accessible.
Our attack is simple and straightforward.
We leverage the shadow model training techniques to generate training data for the attack model which is essentially an image classifier.
Extensive evaluation on three benchmark datasets shows that our proposed attack can effectively infer the architecture/hyperparameters of image classifiers based on convolutional neural network (CNN) given the scientific plot generated from it.
We also reveal that the attack's success is mainly caused by the shape of the scientific plots, and further demonstrate that the attacks are robust in various scenarios.
Given the simplicity and effectiveness of the attack method, our study indicates scientific plots indeed constitute a valid side channel for model information stealing attacks.
To mitigate the attacks, we propose several defense mechanisms that can reduce the original attacks' accuracy while maintaining the plot utility.
However, such defenses can still be bypassed by adaptive attacks.\footnote{Our code is available at \url{https://github.com/boz083/Plot_Steal}.}

\end{abstract}

\section{Introduction}
\label{section:introduction}

Machine learning (ML) has made tremendous progress in various domains during the past decade.
While proven powerful, state-of-the-art ML models require expert knowledge for architecture design.
Also, model developers often need to perform many trials on the hyperparameters to obtain the best performing model.
This process can be quite costly.
Thus, an ML model's information, such as its architecture and hyperparameters, is deemed an important asset of the model owner and must be kept confidential.

\begin{figure}[t]
\centering
\begin{subfigure}{0.118\textwidth}
\centering
\includegraphics[width=\textwidth]{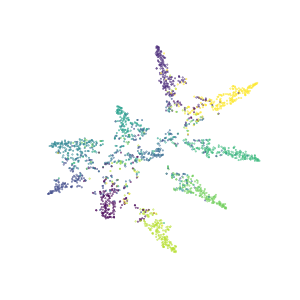}
\caption{ResNet18}
\label{figure:resnet18_intro}
\end{subfigure}\hfill
\begin{subfigure}{0.118\textwidth}
\centering
\includegraphics[width=\textwidth]{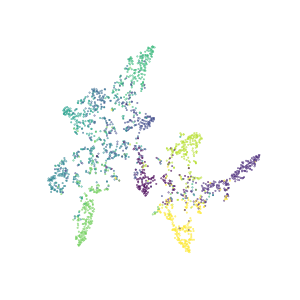}
\caption{ResNet34}
\label{figure:resnet34_intro}
\end{subfigure}\hfill
\begin{subfigure}{0.118\textwidth}
\centering
\includegraphics[width=\textwidth]{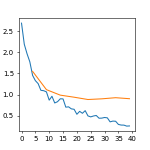}
\caption{ResNet18}
\label{figure:resnet18_intro_loss}
\end{subfigure}\hfill
\begin{subfigure}{0.118\textwidth}
\centering
\includegraphics[width=\textwidth]{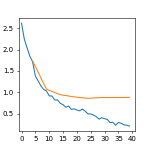}
\caption{ResNet34}
\label{figure:resnet34_intro_loss}
\end{subfigure}
\caption{Examples of t-SNE and loss plots of ResNet18 and ResNet34 models trained on CIFAR-10. 
Scientific plots from different variants of ResNet models indeed show different patterns, which can be exploited by the attackers.}
\label{figure:plots examples}
\end{figure}

Recent studies demonstrate that ML models are vulnerable to information stealing attacks, such as model type~\cite{CSWZ22,LZLYL21} and hyperparameters~\cite{OASF18,WG18}.
These attacks first leverage a dataset to query a target ML model and obtain the responses.
The query-response pairs are then exploited to train an attack model whereby the goal is to infer the information of the target ML model.
To mitigate these attacks, many defenses have been proposed to perturb the information contained in the model's responses or alert the model owner of suspicious queries~\cite{KMAM18,OSF20,KQ20,SHHZ22}.

On the other hand, ML models' scientific plots are easily accessible, via models' project websites or the corresponding research papers/blogs.
For instance, ML model owners often use t-distributed stochastic neighbor embedding (t-SNE)~\cite{MH08} to visualize high-dimensional embeddings generated from their ML models to better understand and interpret model performance.
Loss plots (learning curves) are frequently used during model training to guide model design and debugging (e.g., how fast the model converges, if the learning process is stable, etc.).
Essentially, these scientific plots are abstractions of the model and may directly contain the model's confidential information.
However, to the best of our knowledge, no one has investigated whether scientific plots can be a valid side channel for an adversary to exploit and infer a target ML model's proprietary information.

In this paper, we propose the first \textit{model information stealing} attack that leverages \textit{scientific plots} to steal a target ML model's information, including model type (e.g., ResNet18, ResNet34, or MobileNetV2), training optimizer, training batch size, etc.
The primary contribution of our attack is to show that scientific plots can be a valid side channel to leak the model's proprietary information.
We concentrate on the popular convolutional neural network (CNN) models for image classification and the two most widely used scientific plots in machine learning, i.e., t-SNE and loss plots, shown in \autoref{figure:plots examples}.

\mypara{Attack Methodology}
Given a scientific plot, our goal is to infer information about the model.
We leverage the shadow training technique~\cite{SSSS17,OASF18} to generate a diverse set of data samples, and from that, we train a simple classifier as the attack model.
Concretely, we first prepare shadow models trained with different model types and hyperparameters.
We then generate a set of scientific plots for each shadow model, and label each plot with the shadow model's information.
To train the attack model, we take the $\langle$plot, label$\rangle$ pairs as the training data, and train a Convolution Neural Network (CNN) as the attack model.
Once the attack model is trained, we can infer the information of a model from its scientific plot.

\mypara{Evaluation}
Our evaluation on three datasets, including CIFAR-10, FashionMNIST, and SVHN, shows that the proposed attack can achieve high accuracy.
For instance, on CIFAR-10, given the t-SNE and loss plot generated from a specific model, the attack accuracy for predicting the model's type from a predefined set of 6 popular models is 92.8\% and 95.3\%, respectively.
Given the simple attack method we use, our results demonstrate the severe risk of leaking the model's information by sharing the scientific plots.
Also, we conduct extensive ablation study to show our attack is robust against different plot generation settings (e.g., different density/perplexity for t-SNE plot and with/without axis for loss plots).
We empirically show that our attack performs comparably to existing query-based hyperparameter stealing attacks, yet our attack does not require interaction with the target model.
To reason the success behind our attacks, we further apply Grad-CAM~\cite{SCDVPB17} on our attack models and show that the shape of the t-SNE and the turning point on the loss curve serve as strong signals of the success of the attack.

\mypara{Defense}
To mitigate the attacks, we investigate several defense mechanisms.
We first observe that the defense can be performed under different phases in generating the t-SNE plots, i.e., the embeddings before running t-SNE and the coordinates after running t-SNE.
Also, different perturbation strategies can be applied, including thresholding (saving only the largest $k\%$ embedding values), rounding (saving the values to $k$-th decimal), and noising (adding Gaussian noise to all values).
We find that embedding thresholding before running t-SNE and noising after t-SNE are two effective defense mechanisms.
They can reduce the original attack performance to a large extent while preserving the plot's utility (measured by $k$NN accuracy following~\cite{MH08}).
For loss plots, we find that the sliding window technique can serve as a good defense strategy since it maintains the plot's utility (measured by the average $L_2$ distance of the losses) while largely mitigating the attack performance.
Interestingly, given those defenses, we further show that with proper modification, our attacks can still be effective.
Based on our evaluation, we appeal that scientific plots should be properly perturbed before being published to protect certain proprietary model information.

In summary, we make the following contributions:
\begin{itemize}[noitemsep,topsep=0pt]
\item We propose the first model information stealing attack via scientific plots.
Our evaluation reveals that the attack is effective and robust under different settings.
\item We investigate the success of our attack with the help of Grad-CAM and discover that the attack model captures the essential information from the shape of the plot.
\item We propose several effective defenses to mitigate our attack.
However, we also reveal that an adaptive attacker can bypass the defenses.
\end{itemize}

\section{Preliminary}
\label{section:preliminary}

\mypara{Scientific Plots}
Showing scientific plots is a common way to corroborate the efficacy of ML models.
We briefly introduce two widely used scientific plots, t-SNE and loss plots, which are regularly employed to better understand and visually interpret an ML model's performance.

\noindent\textit{t-SNE Plot.} 
One popular practice to demonstrate an ML model's representation ability is to project some samples' embeddings obtained from the model into the low-dimensional (usually 2-D) space using the t-distributed stochastic neighbor embedding (t-SNE) technique~\cite{MH08}.
In t-SNE, similar embeddings are mapped into nearby places and dissimilar embeddings are projected far away (see \autoref{figure:tsne 6 model examples} for sample plots).
Thus, by observing whether data points from different classes are well separated, we can get a good understanding of the ML model's performance.
We illustrate the detailed procedure in \autoref{section:introduction2tsne}.

\noindent\textit{Loss Plot.}
A loss plot shows the training/validation loss values during the training procedure.
The training loss indicates how well the model fits the training data, while the validation loss indicates how well the model generalizes to validation data that is not used to train the model.
It is a practical way to illustrate the model's generalization ability and convergence rate (see \autoref{figure:loss plot defense} for sample plots).

\mypara{Model Information Stealing Attacks}
Model information stealing attacks aim to infer the type~\cite{LZLYL21,CSWZ22} or hyperparameters of a target model~\cite{OASF18,WG18}.
While existing attacks rely on query responses from the target model to infer model information, we propose a new attack leveraging only the publicly accessible scientific plots.
Our results show that the adversary can successfully infer the detailed hyperparameters of the ML models from these plots (see \autoref{section:evaluation results}).

\section{Threat Model and Methodology}
\label{section:threat_model_and_methodology}

\subsection{Threat Model}
\label{subsection:threat model}

\mypara{Adversary's Goal}
The primary goal of an adversary is to infer key hyperparameters of a target CNN image classifier from its scientific plots.
The inference targets examined in this paper include a selection of popular model types/architectures, optimization algorithms, and batch size settings (see \autoref{subsection:inference targets for tsne attacks} for the complete set of targets considered in this work).
The reason we focus on these targets as they are popular and have been used in a large number of models.
Our attack can certainly incorporate other inference targets as well (see \autoref{section:limiation} for some discussion).
Note that model type/hyperparameter stealing is well recognized by the scientific community~\cite{LZLYL21,CSWZ22,WG18,OASF18}.

\mypara{Adversary's Background Knowledge}
We assume that the adversary has direct access to the scientific plots, for example, a screenshot of plots from PDF or websites.
Although in many scenarios, the adversary might obtain more information from the plot (e.g., high-resolution images, raw data points, vectorized plots), we use screenshots for high accessibility.
The adversary can make adjustments to the plots (e.g., using simple image editing software), including removing axes, labels, plot titles, adjusting color settings, etc.
Moreover, similar to previous works on hyperparameter stealing~\cite{WG18,OASF18}, we assume the adversary has knowledge of the distribution of the target model's training dataset and a selection of candidates for each inference target.

The adversary does not know the data used to generate those plots: For t-SNE plots, the attacker does not know which samples are used for plotting; and in loss plots, the attacker does not know the training/testing samples used to compute the losses.
Besides, the adversary has no query access to the target model (which is different from previous query-based stealing attacks~\cite{WG18,OASF18}).

\mypara{Attack Scenarios}
Sharing scientific plots is common but the associated risk is not well understood.
We believe it is important to systematically evaluate the attack.
Below, we list five realistic scenarios to motivate our study.

\begin{itemize}[noitemsep,topsep=0pt]
\item {The first scenario} is inferring proprietary model information for training a model without tuning architecture or hyperparameters.
\item Alternatively, {the second scenario} is assisting a company to verify if their proprietary models are infringed by the competitors (e.g., by inferring the competitor's model hyperparameters) in a non-intrusive manner (i.e., via the scientific plots published by the competitors).
\item {The third scenario} is serving as an auditing tool to verify the claims in research papers.
We acknowledge that models' information is often specified together with scientific plots in research papers.
However, the information might be incomplete, e.g., batch size and optimization algorithm used are not stated in~\cite{HO21,KJRSL21}.
Also, authors of a considerable portion of papers do not publish their models.\footnote{From our rudimentary search for papers published in IEEE S\&P, CCS, USENIX Security, and NDSS since 2017 that are related to machine learning and have GitHub repositories for their codes, we find 37 papers in total, but only 13 repositories include model weights and hyperparameters of any kind.}
\item Furthermore, the model information obtained by our attack can be leveraged to assist other types of attacks.
As such, {the fourth scenario} is training better surrogate models for generating adversarial examples on a black-box model (see \autoref{subsection:downstream attack}) using the inferred model information from our attack.
\item In the same spirit, {the fifth scenario} is facilitating adversarial reconnaissance to determine potential attacks' difficulty.
For instance, our attack infers the model type, which helps determine whether to launch membership inference attacks against the model (since certain models tend to overfit more than others)~\cite{HSSDYZ21}.
\end{itemize}

\subsection{Attack Methodology}
\label{subsection:attack methodology}

The attack procedure is divided into three steps: shadow model training, scientific plot generation, and attack model training.
We first use shadow models with different model configurations to mimic the behavior of the target models.
Then, those shadow models can be used to generate scientific plots with different model information and train the attack model.

\mypara{Shadow Model Training}
To better capture the characteristics of the target model's information, it is necessary to generate a diverse set of shadow models that are initialized with different model information including model type, optimizer, batch size, etc.
We assume the adversary has a selection of possible values for the inference targets.
Thus, the shadow models are trained with settings randomly selected from the pool.
We follow~\cite{SSSS17,SZHBFB19} and adopt a shadow dataset that comes from the same distribution of the target model's training dataset to train the shadow model.
We later examine the attack with out-of-distribution datasets.
The shadow dataset and the target dataset have no overlap.
Our shadow model training is in line with the latest research direction~\cite{CCNSTT22} whereby many shadow models are trained to attain the attack goal.

\mypara{Scientific Plot Generation}
Once the shadow models are trained, for each shadow model, we can generate a scientific plot.
In this paper, the main example of scientific plots is scatter plots of data points visualized with t-SNE.
Note that we also evaluate the model information leakage via the loss plot where the average training and testing losses are visualized during each training epoch.
Using the trained shadow models, the adversary generates plots with the same setting as the observed one from the target model.

\mypara{Attack Model Training}
The attack model is an image classifier where the input is the generated scientific plots and the output is the corresponding model information such as model type, optimization algorithm, etc.
We train the attack model using the scientific plots generated by the shadow models.
The ground truth labels are the shadow models' information.
Once the attack model is trained, given a scientific plot generated from a specific target model, the attack model can predict its model information.

\mypara{Comparison with Existing Hyperparameter Stealing Attacks}
Model hyperparameter stealing attacks aim to infer the target model's hyperparameters~\cite{OASF18,WG18}.
Typically, they assume the adversary has black-box access (otherwise the problem is trivial) to the target model $f$.
To conduct the attack, the adversary queries $f$ using a query dataset and gets the responses (e.g., predicted probabilities or just labels) from the target model.
By observing the query-response pairs, the attacker then constructs an attack model to infer the hyperparameters.
Existing attacks have an important assumption that the adversary has the (black-box) query access to the target model~\cite{OASF18,WG18}, which means they can leverage an adversarially crafted dataset to query the target model and obtain the response.
Our model information stealing attack does not require any interaction with the target model but only leverages a single publicly accessible scientific plot.
Also, in the scientific plot, the information is compressed.
For example, in the t-SNE plots, the embeddings are projected into only two dimensions using t-SNE and only the average losses instead of the individual losses for training and testing data are reported in the loss plot, which further increases the attack difficulty.
Our evaluation reveals that even in this case, our proposed attack, albeit simple and straightforward, can still effectively infer the model information.

\section{Evaluation Setup}
\label{section:evaluation setup}

In this section, we describe the default experiment settings.
Later we also conduct a series of ablation studies to show our attack is robust in different settings in \autoref{subsection:ablation study}.

\mypara{Shadow and Target Model Training}
We use three benchmark datasets, CIFAR-10, FashionMNIST, and SVHN, in the experiments.
Each dataset is divided into 4 non-overlapping partitions, namely shadow training, shadow testing, target training, and target testing.
For each shadow/target model, we randomly sample 20,000 data points for training.

\begin{figure*}[!t]
\begin{subfigure}{0.33\textwidth}
\centering
\includegraphics[width=\textwidth]{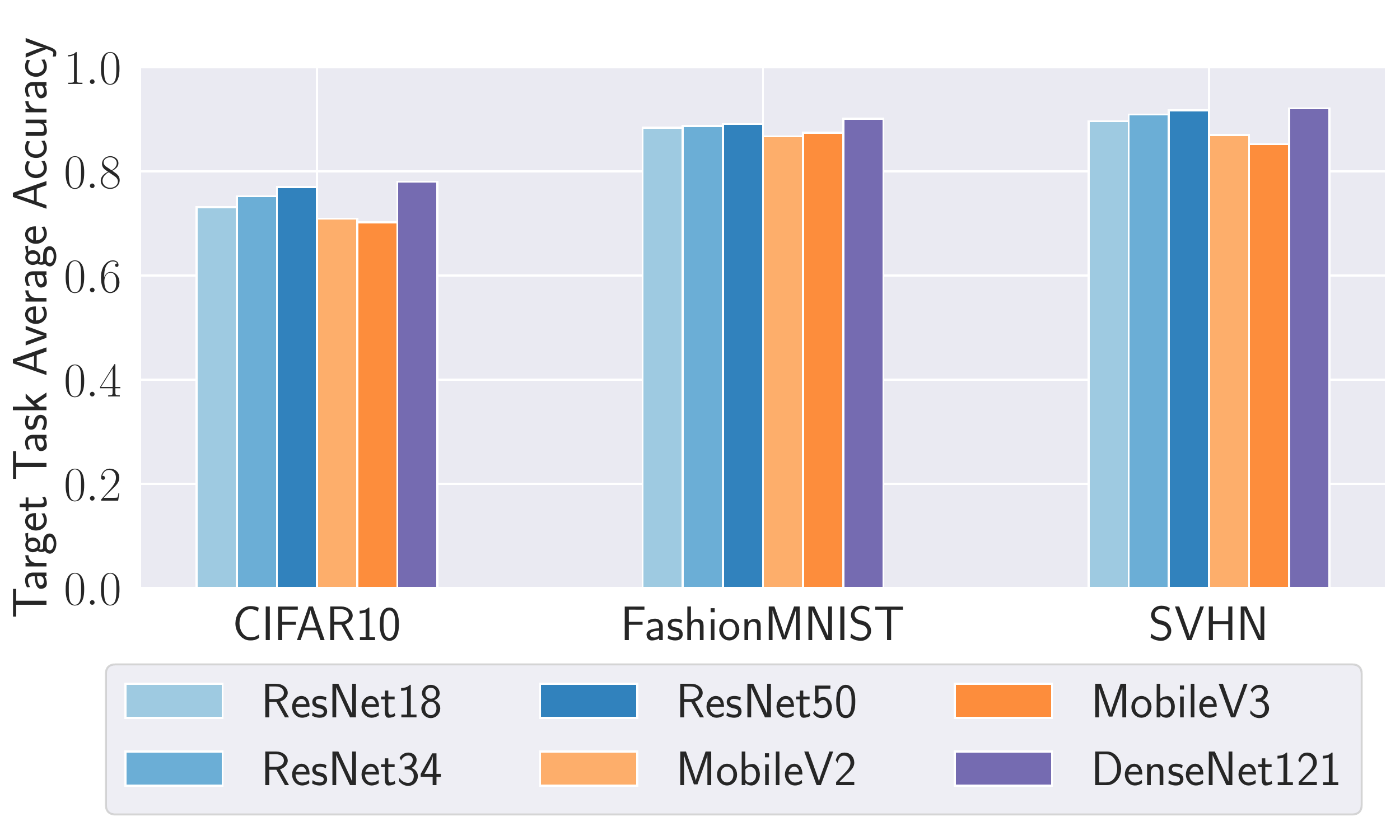}
\caption{Model Types}
\label{figure:target performance model}
\end{subfigure}%
\hfill
\begin{subfigure}{0.33\textwidth}
\centering
\includegraphics[width=\textwidth]{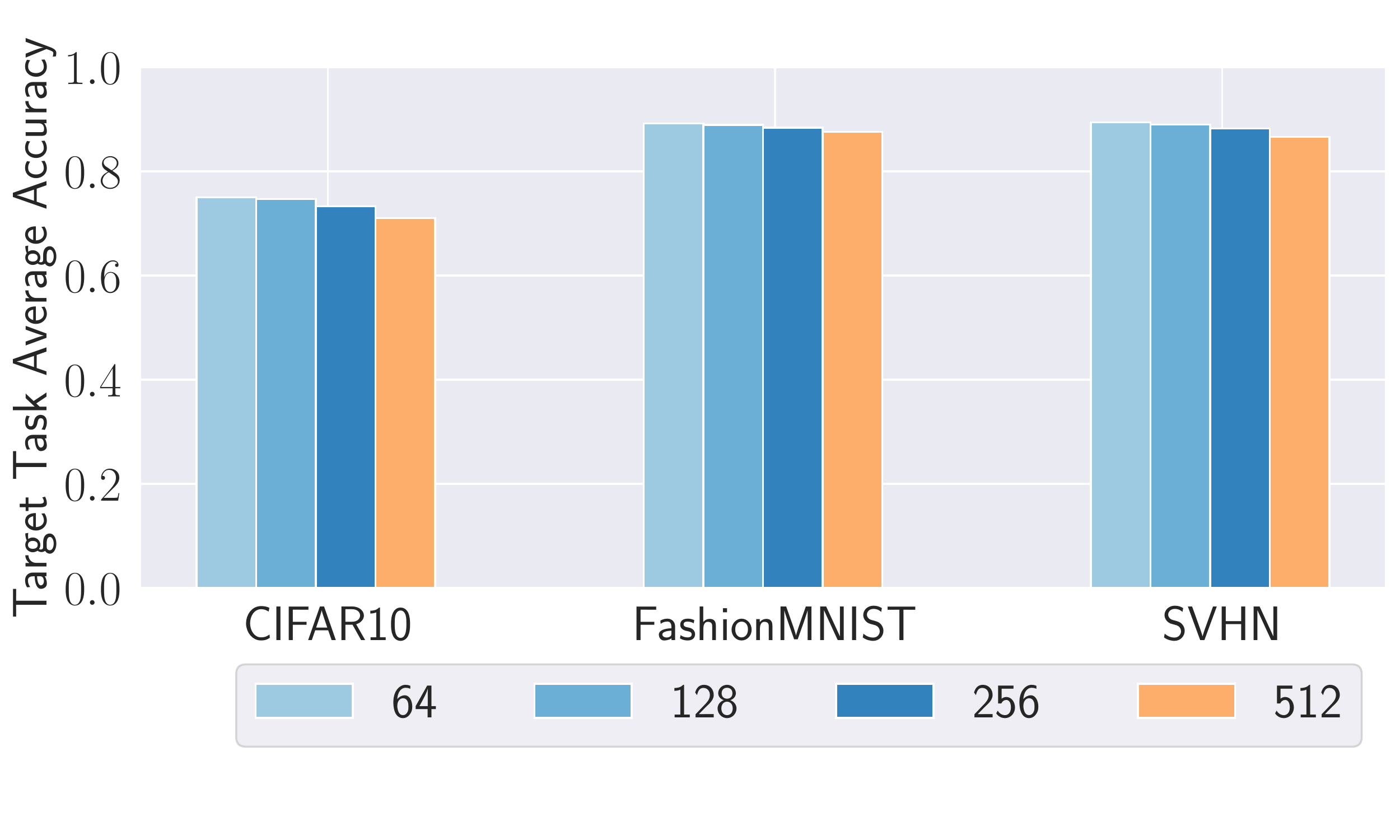}
\caption{Batch Size}
\label{figure:target performance bs}
\end{subfigure}%
\hfill
\begin{subfigure}{0.33\textwidth}
\centering
\includegraphics[width=\textwidth]{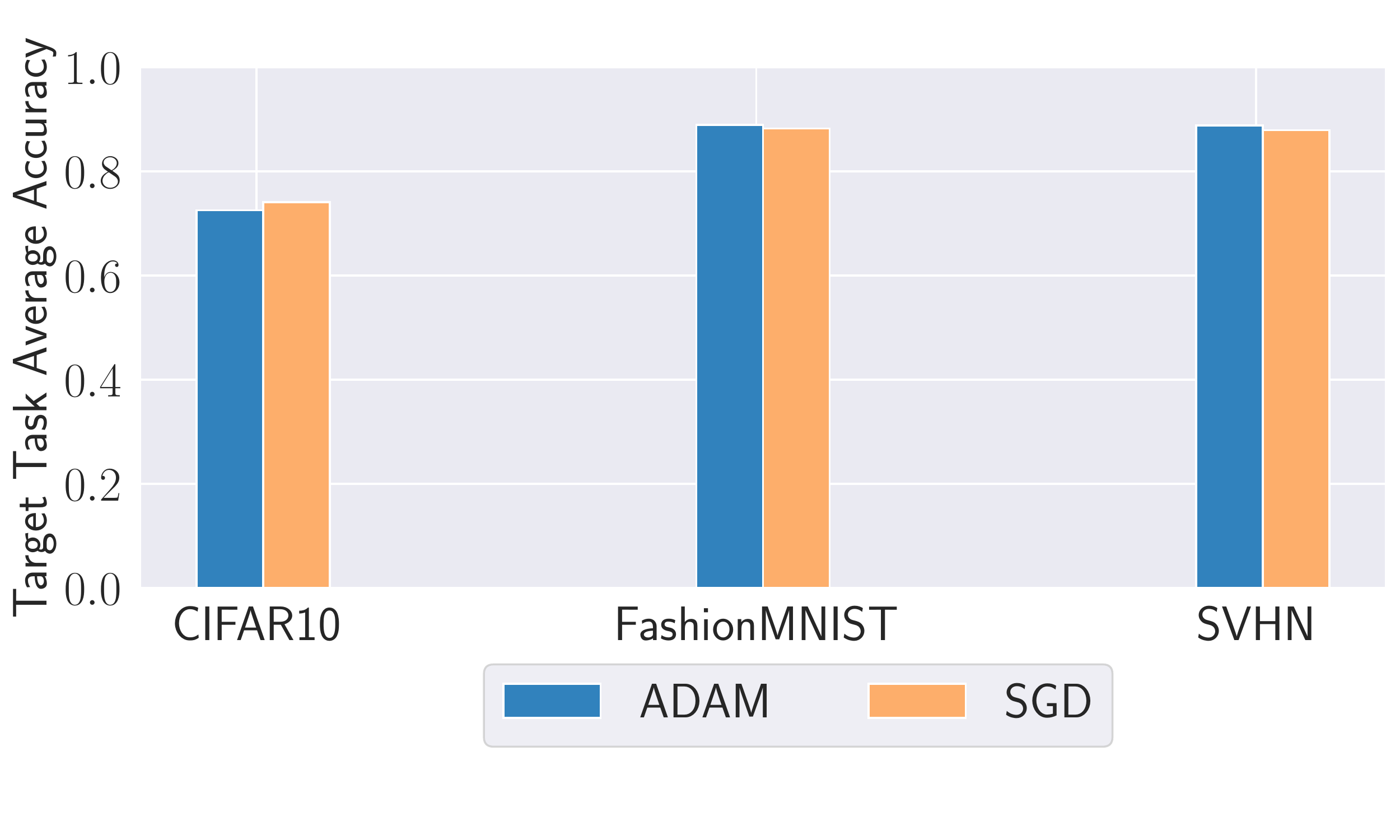}
\caption{Optimization Algorithms}
\label{figure:target performance optim}
\end{subfigure}%
\caption{Average accuracy on the original classification tasks for models trained on 3 different datasets. 
The target task performances remain similar across different model types, batch size and optimization algorithms.}
\label{figure:target task performance}
\end{figure*}

We also use the popular approach of fine-tuning pre-trained models  ~\cite{ZYMK16,CKNH20,HFWXG20} and adopt the models pre-trained from ImageNet (if available) as initialization for further training.
The choice of architecture and training hyperparameters are randomly sampled from the pool of possible values.
In total 2250 shadow models and 750 target models are trained for each dataset.
To ensure all shadow and target models are properly trained, we discard low-performing models (test accuracy below 50\% on the target task).
The trained shadow and target models have relatively close performance on the target task, as seen in \autoref{figure:target task performance}.

\mypara{t-SNE Creation}
In the default setting, for each trained shadow/target model, we randomly select 2,000 samples from the corresponding test dataset.
We then follow the widely-used settings to generate t-SNE plots.
We query the model with these samples and take the output of the second to last layer as the samples' embeddings to generate the t-SNE plot.
The plots are saved as images without axis, labels, and titles, keeping only the scattered sample points in 2-dimensional space.
Different colors are used in t-SNE plots to denote those samples' classes in the original classification task.
However, to extend the range of possible t-SNE plots used for target models, we convert the color t-SNE plots into grayscale t-SNE plots.
This eliminates the chance that the attack relies on difference in color schemes.
We later observe that the attack performance remains unchanged for color, grayscale, or binary (i.e., 1-bit monochrome) t-SNE plots (see \autoref{table:color ablation}).
The t-SNE plots used for experiments are 300x300 PNG images with 100dpi.
Most t-SNE plots used in scientific papers and blog posts tend to have higher dimension/definition~\cite{tsne-blog-g,tsne-blog-joe,tsne-blog-or}.

\mypara{Loss Plot Creation}
During the shadow model training, we record the average training loss 5 times per epoch and the average test loss every epoch.
Both training and testing loss curves are then plotted with different colors.
Normally, loss plots have axis information to denote the training epoch and loss value.
To investigate whether such information facilitates the attack, we generate two types of loss plots, i.e., with or without axis information.

\mypara{Attack Model Training}
For fast convergence, we leverage ResNet18~\cite{HZRS16} pre-trained on ImageNet as the base attack model.
For t-SNE plots, we fine-tune the attack model for 80 epochs with batch size 32, learning rate 0.0001, and Adam as its optimizer.
For loss plots, we fine-tune the attack model for 40 epochs, and the rest of the settings are the same as above.

\mypara{Runtime Configuration}
The experiments are repeated 10 times.
We report the mean as well as standard deviation values.
For each run, we follow the same experimental setup.

\section{t-SNE Evaluation}
\label{section:evaluation results}

\subsection{Model Information Stealing Attacks Against t-SNE}
\label{subsection:inference targets for tsne attacks}

In this section, we first highlight what model information can be inferred from t-SNE plots.
Here we consider three different inference targets:
(1) the target model's \textit{model type}, (2) the \textit{optimization algorithm}, and (3) the \textit{batch size} used in the target model's training procedure.
The attack is evaluated under both the mixed and fixed settings.
The mixed setting is the default setting described in \autoref{section:evaluation setup}.
In fixed setting, the adversary has other knowledge of the model information (e.g., the adversary knows the model type and batch size when inferring optimization algorithm).
We also investigate our attack performance on additional inference targets within \textit{model architectures} by building custom models.

\begin{figure*}[!t]
\centering
\begin{subfigure}{0.16\textwidth}
\centering
\includegraphics[width=\textwidth]{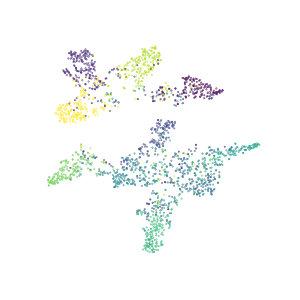}
\caption{ResNet18}
\label{figure:tsne resnet18}
\end{subfigure}%
\begin{subfigure}{0.16\textwidth}
\centering
\includegraphics[width=\textwidth]{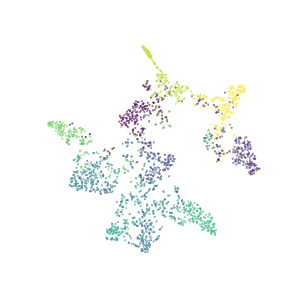}
\caption{ResNet34}
\label{figure:tsne resnet34}
\end{subfigure}%
\begin{subfigure}{0.16\textwidth}
\centering
\includegraphics[width=\textwidth]{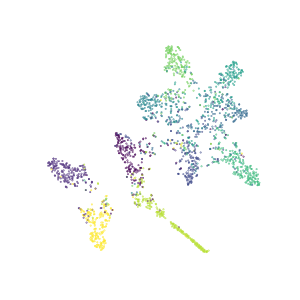}
\caption{ResNet50}
\label{figure:tsne resnet50}
\end{subfigure}%
\begin{subfigure}{0.16\textwidth}
\centering
\includegraphics[width=\textwidth]{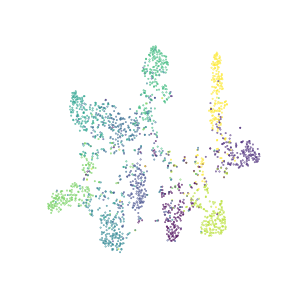}
\caption{MobileV2}
\label{figure:tsne mobilev2}
\end{subfigure}%
\begin{subfigure}{0.16\textwidth}
\centering
\includegraphics[width=\textwidth]{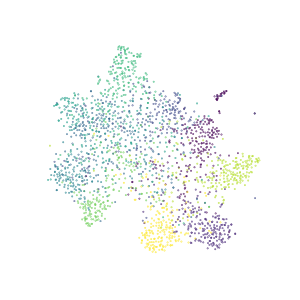}
\caption{MobileV3}
\label{figure:tsne mobilev3}
\end{subfigure}%
\begin{subfigure}{0.16\textwidth}
\centering
\includegraphics[width=\textwidth]{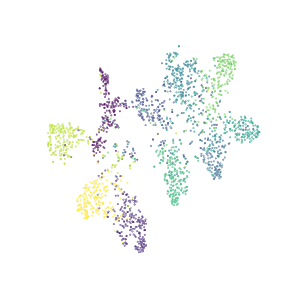}
\caption{DenseNet121}
\label{figure:tsne densenet}
\end{subfigure}%
\caption{t-SNE plots generated from models of different types (batch size 128, Adam Optimizer).}
\label{figure:tsne 6 model examples}
\end{figure*}  

\mypara{1. Model Type Inference}
We consider six popular model types: ResNet18, ResNet34, ResNet50, MobileNetV2, MobileNetV3, and DenseNet121.
They belong to three widely used model families, i.e., ResNet~\cite{HZRS16}, MobileNet~\cite{HZCKWWAA17}, and DenseNet~\cite{HLMW17}.
\emph{We intentionally selects different types of models from the same family to increase the attack difficulty, since models from the same family behave similarly and are harder to distinguish.}
We first conduct our attack by inferring the target model’s family and then the more fine-grained actual model type.

From \autoref{table:model inference performance}, we observe the attack model can extract the model family and more fine-grained model type information for all three datasets (the first and second rows).
For instance, on CIFAR-10, the model family (type) prediction reaches 85.7\% (77.1\%) accuracy.
The confusion matrix of the inference results (\autoref{figure:confusion cifar10} in Appendix) shows that the attack model can accurately identify model types even within the same family.
The values in the parenthesis show the attack performance in a fixed setting.
In this case, the attack performs exceedingly well, with inference accuracy at around 90\% on model types for all 3 datasets.
Although this is a less realistic attack scenario, we use it to showcase the strength of this side-channel attack when the adversary has additional knowledge.

From a more intuitive perspective, \autoref{figure:tsne 6 model examples} shows the example t-SNE plots generated from different model types, which indeed have different patterns.
For example, ResNet family's t-SNEs show multiple sharp edges in clusters, while those of the MobileNet family have more rounded clusters.

\begin{table}[!t]
\centering
\caption{Average attack accuracy for different inference targets on 3 different datasets. 
The number of candidates for inference targets are in brackets. 
The values in parenthesis are results in fixed setting.}
\label{table:model inference performance}
\tabcolsep 2pt
\scalebox{0.7}{
\begin{tabular}{l c c c c} 
\toprule
\textbf{Inference Targets} & \textbf{CIFAR-10} & \textbf{FashionMNIST} & \textbf{SVHN} \\ 
\midrule
Model Family <3> & $85.7(94.4) \pm 0.6\%$ & $84.8(98.9) \pm 1.0\%$ & $81.0(96.9) \pm 0.9\%$ \\ 
Model Type <6> & $77.1(92.8) \pm 1.0\%$ & $75.1(88.0) \pm 1.3\%$ & $74.1(95.2) \pm 1.0\%$ \\
Optimizer <2> & $96.1(100.0) \pm 0.2\%$ & $72.9(99.8) \pm 1.8\%$ & $97.4(100.0) \pm 0.2\%$ \\
Batch Size <3> & $69.4(90.3) \pm 1.4\%$ & $65.7(96.3) \pm 0.6\%$ & $66.2(84.8) \pm 0.5\%$ \\
Batch Size <4> & $60.5(77.4) \pm 0.7\%$ & $52.8(93.6) \pm 0.7\%$ & $57.4(74.4) \pm 0.9\%$ \\
\bottomrule
\end{tabular}
}
\end{table} 

\mypara{2. Optimization Algorithm}
The optimization algorithms are also a crucial part of training the ML model.
In our evaluation, we consider two commonly used optimization algorithms, Adam and SGD.
Two sets of target models are trained to have similar prediction performance.
The average accuracy on the CIFAR-10 classification task is 72.5\%/74.1\% for models optimized with Adam/SGD and both have $\le1$\% standard deviation.
The difference in average accuracy for the other 2 datasets are both below 1\%.

Interestingly, although the models trained with different optimization algorithms have similar performance, the t-SNE plots are significantly different (see \autoref{figure:optimization tsne} in Appendix).
With such differences, the attack model achieves close to 100\% accuracy (less than 1\% from perfect prediction in fixed setting for all 3 datasets) in inferring the optimization algorithms on both CIFAR-10 and SVHN datasets.
We suspect the difference is due to the different ratio of 0 values in the embeddings, given the fact that dimension reduction algorithms like t-SNE typically are sensitive to input sparsity~\cite{MH08}.
For instance, on CIFAR-10, we find that the embeddings 52.6\% 0 values if optimized by SGD while 68.6\% for Adam.
The significant difference in t-SNE plots cannot be generalized to FashionMNIST dataset, even though the attack still achieves inference accuracy higher than random guessing.
The embeddings from models optimized by the two algorithms also have similar sparsity, as conjectured (FashionMNIST ResNet18 models have 52.0\% and 49.4\% 0 values when optimized by SGD and Adam respectively.).

\mypara{3. Batch Size}
The last two rows of \autoref{table:model inference performance} show that the attack model can successfully infer batch size information (64 vs 128 vs 256 vs 512) of the target models from generated t-SNE plots.
The attack performance is lower compared to model types and optimization algorithms from previous sections.
However, the inference accuracy is still much higher than the random guessing baseline for all 3 datasets.
For instance, when the potential batch sizes are 64 vs 128 vs 512, our attack can reach 69.4\% accuracy on CIFAR-10.
Similar batch sizes are more easily misclassified (see \autoref{figure:confusion bs} in Appendix).
When a strong adversary has knowledge of model type and optimization algorithm, the attack performance also improves significantly.
The 3-class batch size inference on all 3 datasets have attack accuracy higher than 80\% in this setting, notably reaching 96.3\% on FashionMNIST dataset.

\begin{table}[!t]
\centering
\caption{The average attack accuracy for different inference targets on custom models.}
\label{table:custom model inference performance}
\scalebox{0.7}{
\begin{tabular}{l c c c c} 
\toprule
\textbf{Inference Targets} & \textbf{Possible Values} & \textbf{Attack Accuracy} \\ 
\midrule
Activation Function & relu, elu, tanh & $92.4 \pm 0.6\%$ \\
\#.\ FC.\ Layers & 2, 3, 4 & $81.9 \pm 1.3\%$  \\
\#.\ CONV.\ Layers & 2, 3, 4 & $63.9 \pm 1.0\%$ \\ 
\#.\ Kernel Size & 3, 5 & $67.1 \pm 0.6\%$  \\
Dropout & True, False & $54.2 \pm 1.3\%$  \\
Max-pooling & True, False & $61.8 \pm 1.9\%$  \\
Batch Size & 64, 128, 256, 512 & $37.8 \pm 0.8\%$ \\
Optimizer  & Adam, SGD & $70.3 \pm 3.4\%$ \\
\bottomrule
\end{tabular}
}
\end{table} 

\begin{figure}[!t]
\centering
\begin{subfigure}{0.33 \columnwidth}
\centering
\includegraphics[width=\textwidth]{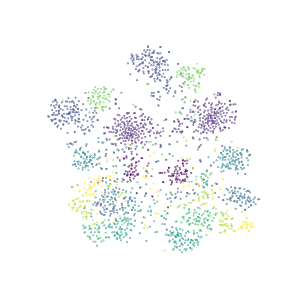}
\caption{conv=4, ks=3,\\
fc=2, bs=256,\\
elu, Adam}
\label{figure:tsne custom cnn1}
\end{subfigure}%
\hfill
\begin{subfigure}{0.33 \columnwidth}
\centering
\includegraphics[width=\textwidth]{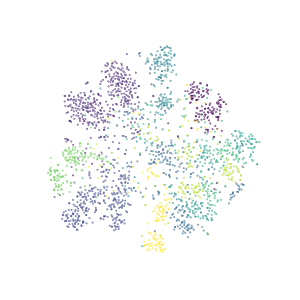}
\caption{conv=3, bs=5,\\ 
fc=2, bs=512,\\ 
relu, SGD}
\label{figure:tsne custom cnn2}
\end{subfigure}%
\hfill
\begin{subfigure}{0.33 \columnwidth}
\centering
\includegraphics[width=\textwidth]{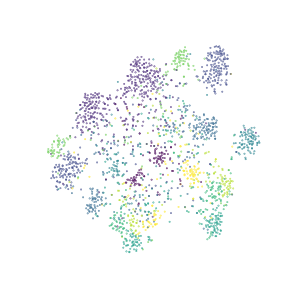}
\caption{conv=2, ks=3,\\
fc=3, bs=128,\\
tanh, Adam}
\label{figure:tsne custom cnn3}
\end{subfigure}
\caption{t-SNE plots of custom CNNs trained on SVHN.}
\label{figure:custom cnn tsne}
\end{figure}

\mypara{4. Custom Model Architecture Inference}
Results from above show our attack model can precisely infer the model type from t-SNE plots.
However, those models with different types still differ notably from each other, e.g., ResNet18 and ResNet34 have 18 and 34 layers, respectively.
We further investigate whether our attack remains effective to the {\it models only with subtle differences}.
The rationale is that the more subtle the difference is, the more difficult for the attacker to materialize the information inference attack.
To this end, we construct custom CNN models based on 6 key hyperparameters, similar to the ones investigated in the previous work.
\autoref{figure:custom cnn tsne} shows examples of t-SNEs from models with different hyperparameters trained on SVHN.
The t-SNE plots from custom models become almost indistinguishable for humans.
The custom CNN model is comprised of 2-4 convolution layers with a kernel size of 3 or 5 and 64 channels, 2-4 fully connected layers with 512 neurons, and a final fully connected layer that acts as the classifier.
There are optional max-pooling after the convolution layers and optional dropout with a probability of 0.4 after the fully connected layers.
The activation functions used are one of relu, elu, and tanh for each model.
The batch size and optimization algorithms have the same selection pool as in previous sections.
All custom models are trained for 30 epochs and low-performing models are discarded.
A total of 1,795 shadow models and 295 target models are used to generate corresponding t-SNE plots.
The detailed hyperparameter targets, selections of values, as well as inference performance, are shown in \autoref{table:custom model inference performance}.
We observe that the number of convolution layers, number of fully connected layers, and activation functions have especially high information leakage from t-SNE plots.
For instance, the attack model can infer activation used in the custom model with 92.4\%, given a selection pool of relu, elu, and tanh.
The attacks on the 2 inference targets investigated in previous sections, batch size and optimization algorithms, still achieve good performance.
The performance gap to the previous predetermined 6 model types is yet noticeable, due to the significant increase in difficulty (i.e., t-SNE plots are much more similar to one another).
The attack models, however, have attained sub-optimal performance on inferring whether dropout or max-pooling are used and the kernel size of the convolution layers.

\mybox{
Our evaluation shows that t-SNE plots can be a valid side channel to infer the model information.
Among all inference targets, activation function, number of fully connected layers, number of convolutional layers, and the optimization algorithm are more vulnerable than the others.
}

\subsection{Ablation Study}
\label{subsection:ablation study}

We further investigate whether our attack is still effective with: (1) fewer shadow models, (2) different color settings for creating the t-SNE plots, (3) different numbers of sample points for creating t-SNE, and (4) different perplexity settings in t-SNE.
Since different optimization algorithms yield significantly different t-SNEs, we focus on model type inference and batch size inference tasks instead, which are more difficult.
The experiments are conducted on SVHN.

\begin{figure}[!t]
\centering
\includegraphics[width=0.7\columnwidth]{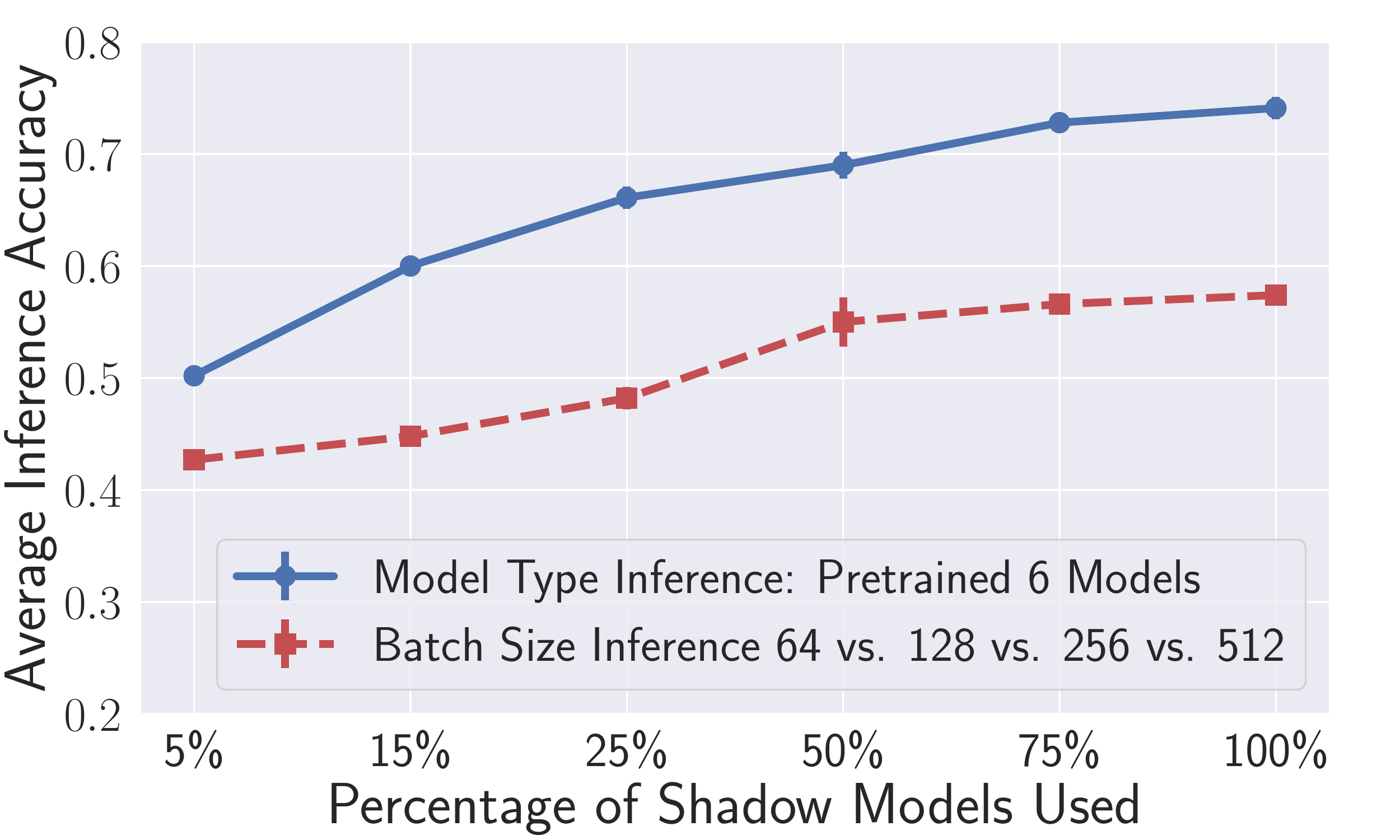}  
\caption{Average inference accuracy by using different numbers of shadow models to train the attack model on SVHN.}
\label{figure:shadow ablation}
\end{figure}

\mypara{Number of Shadow Models}
\autoref{figure:shadow ablation} shows the performance of model type inference and batch size inference when using different percentages of total shadow models trained.
The inference performance indeed increases with more shadow models used for training.
However, even with 5\% of the shadow models, the attack model can already achieve good model type inference accuracy of 50.2\%.
With 50\% shadow models, the inference performance is within 4\% of the default setting's attack accuracy.
We also have similar observations in batch size inference.

\mypara{Color Settings}
We test the inference performance when the attack model is trained on t-SNE plots with different color settings, including the original color setting, grayscale, and binary.
\autoref{figure:tsne color} (in Appendix) shows the comparison of the three color settings.
Originally, different colors denote different classes.
Grayscale makes it harder to differentiate classes, and binary makes classes indistinguishable.
\autoref{table:color ablation} shows the inference accuracy generally remains unchanged with all three t-SNE color settings.
For instance, for model type inference, the attack accuracy is 74.0\%, 74.1\%, and 73.6\% for color, grayscale, and binary t-SNE plots, respectively.
Our evaluation results reveal that the shape instead of the color of the t-SNE plot plays the most important role in distinguishing the hyperparameters.
This makes our attack more practical to the t-SNE plot in the real world.
We use Grad-CAM to provide an in-depth visual explanation in \autoref{section:grad-cam analysis}.

\begin{table}[!t]
\centering
\caption{Inference accuracy of models trained by different color settings' t-SNE plots on SVHN.}
\label{table:color ablation}
\scalebox{0.7}{
\tabcolsep 4pt
\begin{tabular}{l  c c c} 
\toprule
\textbf{Inference Task} & \textbf{Color} & \textbf{GrayScale}  & \textbf{Binary} \\  
\midrule
Model Type  & $74.0 \pm 1.1\%$ & $74.1 \pm 1.0\%$ & $73.6 \pm 1.0\%$ \\ 
Batch Size (64 vs 128 vs 256 vs 512) & $60.6 \pm 0.6\%$ & $57.4 \pm 0.9\%$ & $57.4 \pm 0.8\%$  \\
\bottomrule
\end{tabular}
}
\end{table} 

\begin{figure}[!t]
\centering
\includegraphics[width=0.7\columnwidth]{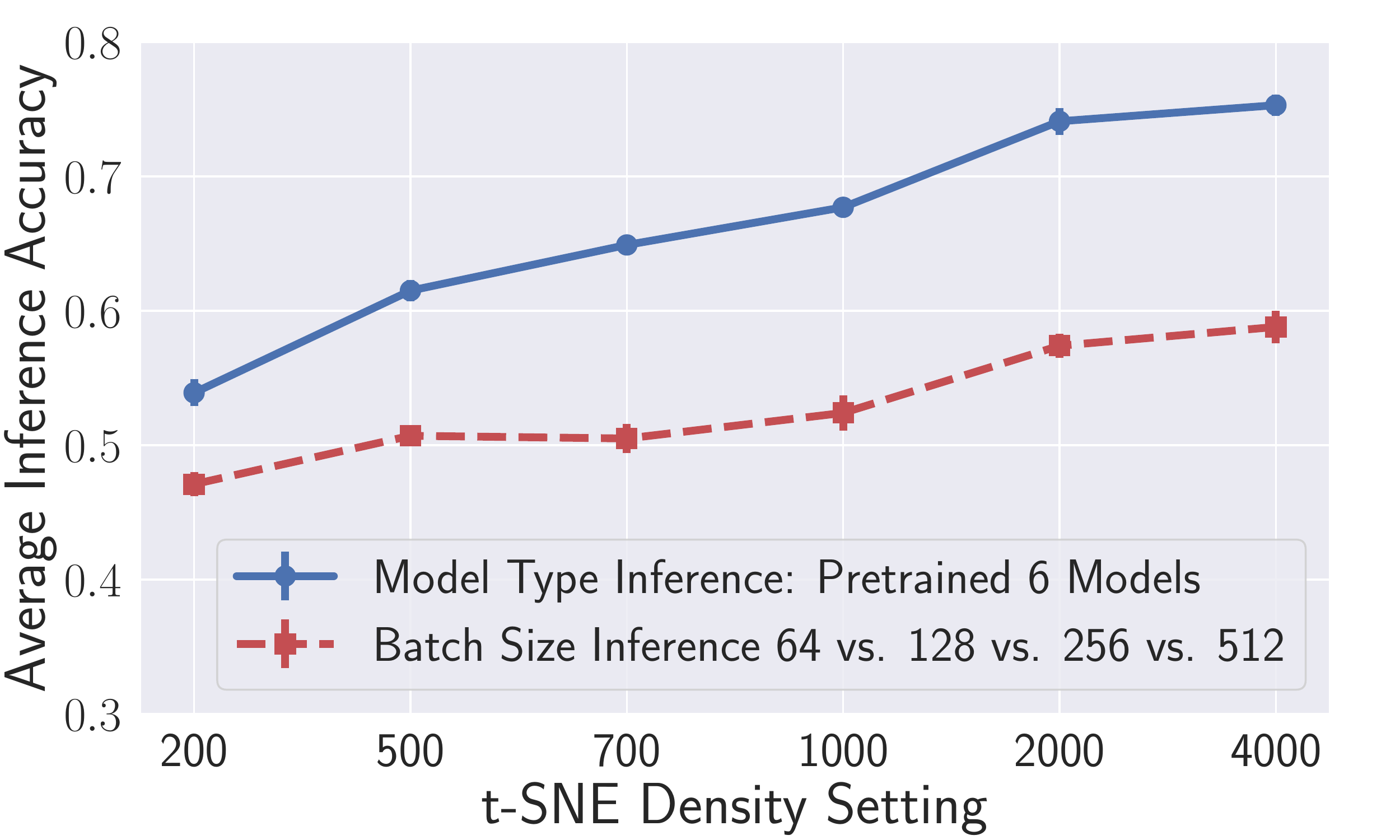}  
\caption{Inference performance at different t-SNE density settings. 
The models are trained on SVHN.}
\label{figure:density ablation}
\end{figure}

\mypara{Density Settings}
Density denotes the number of sample points used to fit t-SNE and make the plot.
As shown in \autoref{figure:tsne density} (in Appendix), the density setting also affects the clusters’ geometric characteristics.
\autoref{figure:density ablation} shows the inference performance increases as density increases.
The inference accuracy of model type is 68.8\% with 1,000 points/plot while only 55.2\% with 200 points/plot, respectively 5.1\% and 18.7\% lower than the benchmark at 2,000 points/plot.
At low-density settings, not enough sample points are used to fit t-SNE that forms clusters with unique information about the target models.
Since the marker size remains unchanged, a lower density setting also results in more empty spaces in the plots.
Once enough points are used, t-SNE plots represent the target models accurately and increasing density does not add more information.

\begin{figure}[!t]
\centering
\includegraphics[width=0.7\columnwidth]{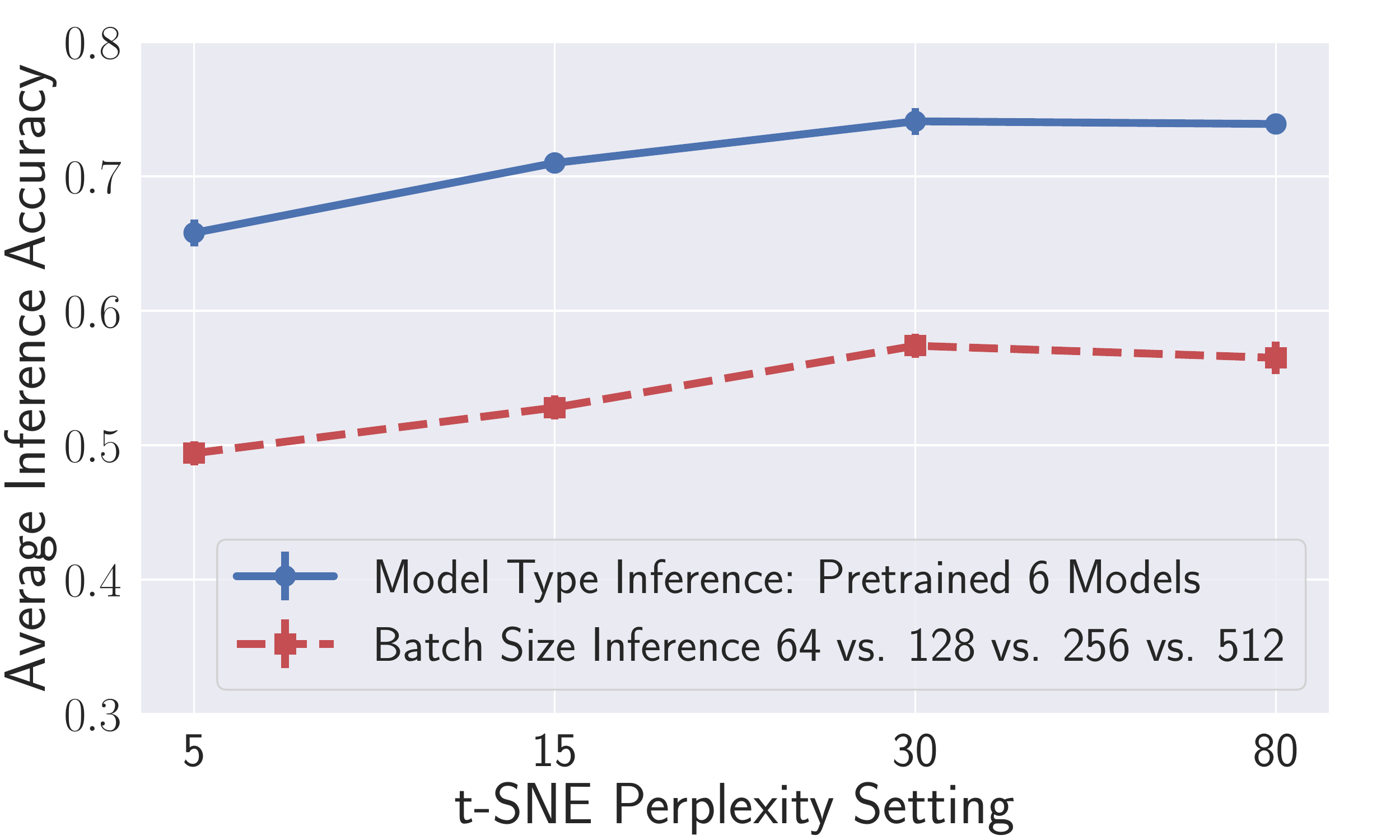}  
\caption{Inference performance with different perplexity settings.
The models are trained on SVHN.}
\label{figure:perplexity ablation}
\end{figure}

\mypara{Perplexity Settings}
Perplexity is an important hyperparameter for generating t-SNE which controls the number of nearest neighbors used for calculating cluster centers during the fitting process of t-SNE.
With a larger number of samples involved (i.e., a larger density), a higher perplexity is preferred.
\autoref{figure:tsne perplexity} (in Appendix) shows the example t-SNEs with different perplexity settings from the same set of embeddings of a ResNet50 model trained on SVHN (we have a similar observation for other model types).
Note that 5 and 80 are the recommended lower and upper bounds of the values, and 30 is the default value~\cite{TsneDefault}.
As we do not observe a clear change between 30-80, we try a smaller value 15 to make the plots more diverse.
From \autoref{figure:perplexity ablation}, we notice that inference performance positively correlates with perplexity.
When the perplexity is 15, the attack accuracy of model type inference increases 5.2\% compared to 5, while the accuracy remains similar when the perplexity increases from 30 to 80.
The increased performance with larger perplexity can be explained by observing t-SNE plotted at different perplexity settings in \autoref{figure:tsne perplexity}.
When perplexity is set too low for the given dataset’s size, the fitted t-SNE does not represent the target model's embeddings properly.
Once the perplexity is high enough, different settings still produce different cluster shapes but share similar geometric characteristics.

\mybox{
Our attack remains effective even with small numbers of shadow models and is not significantly affected by the color, density, and perplexity of t-SNE plots.
}

\subsection{Open-World Settings}
\label{subsection:open-world settings}

To simulate more realistic attack scenarios, we relax the constraints on the attacker’s knowledge of the target model, including dataset distribution and t-SNE settings.
In the open-world setting, the adversary might not have access to datasets with the same distribution.
The exact settings used for t-SNE plots can also be difficult to obtain based on observation.
We demonstrate our attack model’s robustness in the open-world setting using the model type inference task.

\mypara{Mixed Datasets}
The geometric characteristics of t-SNE plots highly depend on the dataset used for the target task.
\autoref{figure:tsne datasets} shows examples of t-SNE plots created with embeddings from target models trained on the three datasets, which have different patterns.
While the attack model performs well for all three datasets separately, as shown above, we evaluate the attack performance with mixed datasets, simulating the scenario where the attacker has a well-trained model including multiple datasets.
As shown in \autoref{table:mixed dataset inference}, when the attack model is trained on all three datasets, the inference accuracy on the mixture of t-SNE plots created on three datasets is 73.5\%, which shows no deterioration.

We also evaluate attack performance for out-of-distribution data, where the shadow models and target models are trained with different datasets, to assess whether the model type information's characteristics in t-SNE are shared across datasets.
The second row in \autoref{table:mixed dataset inference} shows that the inference performance decreases greatly, but is still higher than random guessing (16.7\%), which means the attack model still can extract model information from t-SNE plots generated from different dataset distributions.
Domain shift~\cite{QSSL08} is one of the biggest challenges when deploying machine learning models in the real world.
This certainly applies to our attack models as well.
We discuss this limitation in \autoref{section:limiation}.

To overcome this limitation, we further evaluate whether our attack can generalize to different datasets given only a small fraction of shadow models trained on the new dataset.
Concretely, we fine-tune the attack model trained on CIFAR-10 using only a small number of t-SNE plots generated from shadow models trained on SVHN.
We evaluate model type inference performance on t-SNE plots built from target models trained on SVHN.
\autoref{table:mixed dataset inference} shows that the inference performance increases from 22.9\% to 60.3\% with only 5\% shadow models trained on SVHN added.
Recall that the randomly initialized attack model trained with 5\% shadow models only achieves 50.2\% inference accuracy (10.1\% improvement with fine-tuning).
The benefit of using an attack model pre-trained on out-of-distribution data decreases as the number of shadow models available increases.
With 25\% of the original number of shadow models, fine-tuning on CIFAR-10 attack model offers only 0.3\% increase in attack accuracy, compared to the random initialization.
Our observation reveals that the attack can easily transfer to other dataset distributions, which further demonstrates the severe model information leakage risks stemming from t-SNE.

\begin{table}[!t]
\centering
\caption{Model type inference with mixed datasets.}
\label{table:mixed dataset inference}
\scalebox{0.7}{
\begin{tabular}{l l c c} 
\toprule
\textbf{Training data} & \textbf{Testing Data} & \textbf{Accuracy}\\ 
\midrule
Combined (All 3) & Combined (All 3) & $73.5 \pm 0.2\%$\\
CIFAR10 & SVHN & $22.9 \pm 0.1\%$\\
5\% SVHN (fine-tuned) & SVHN & $60.3 \pm 1.0\%$  \\
15\% SVHN (fine-tuned) & SVHN & $64.9 \pm 1.0\%$  \\
25\% SVHN (fine-tuned) & SVHN & $67.3 \pm 0.5\%$  \\
\bottomrule
\end{tabular}
}
\end{table} 

\begin{figure}[!t]
\centering
\includegraphics[width=0.7\columnwidth]{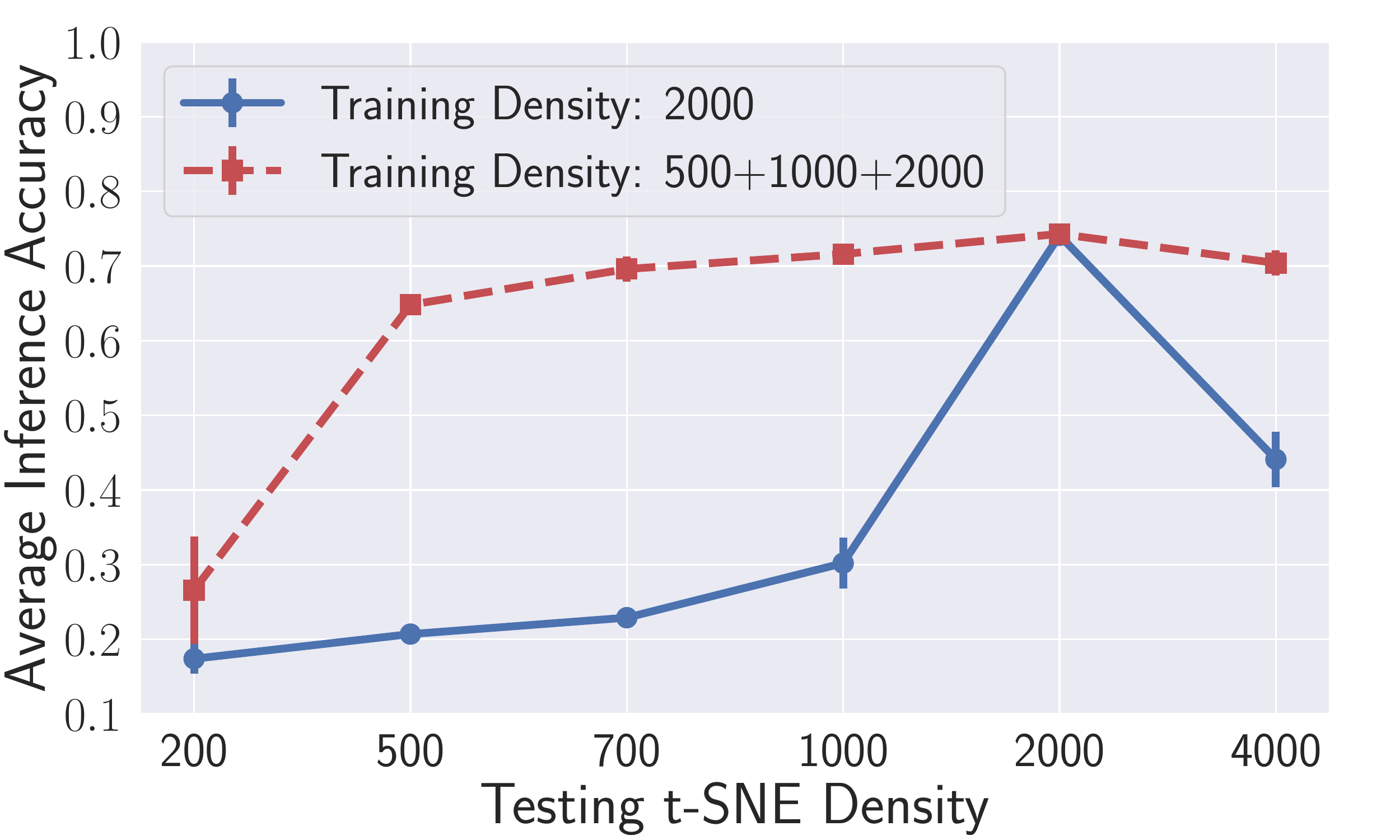}  
\caption{Model type inference performance with different t-SNE sample densities for models trained on SVHN.}
\label{figure:density transfer}
\end{figure}

\mypara{Density Transferability}
Previously, we assume the t-SNE plots of shadow models and the target model share the same densities.
In practice, the densities could be different.
While the attacker could re-train the shadow models with the same densities, we are interested in whether the attack can transfer to different density settings.
Here we consider two different settings where the first one only leverages 2,000 samples to generate the t-SNE plots and the second one leverages 500, 1,000, and 2,000 samples to generate t-SNE plots during the training procedure.
From \autoref{figure:density transfer}, we observe that the attack model trained with 3 different densities can perform better in different density settings even if the testing density is unseen during the training.
For instance, when the testing density is 4,000, the model type inference accuracy is 70.4\% for the attack model trained with 3 different densities while only 44.1\% for the attack model trained with 2,000 density t-SNE plots.
The result demonstrates that the attack can successfully transfer to various density settings by training with t-SNE plots with limited combinations of density settings.

\mypara{Perplexity Transferability} 
In \autoref{subsection:ablation study}, we show perplexity does not significantly affect inference results if it is set high enough.
\autoref{table:perplexity transferability} shows the inference performance with t-SNE built with mixed perplexities.
We observe that the attack model trained with t-SNE plots with single perplexity can generalize to t-SNE plots with different perplexity settings, especially for a larger perplexity.
For instance, when trained with 30 perplexity t-SNE, the testing accuracy is 47.3\% on t-SNE plots with 80 perplexity.
We also find that having mixed perplexity in training improves the attack model’s ability to generalize predictions for t-SNE with unseen perplexity.
Both interpolation and extrapolation of perplexity show good inference performance and are significantly better than the performance of the attack model trained with single perplexity.

\begin{table}[!t]
\centering
\caption{Model type inference with mixed perplexity.
The models are trained on SVHN.}
\label{table:perplexity transferability}
\scalebox{0.7}{
\begin{tabular}{c c c c} 
\toprule
\textbf{Training Perplexity} & \textbf{Testing Perplexity} & \textbf{Accuracy}\\ 
\midrule
30 &  5 & $17.3 \pm 1.9\%$\\ 
30 & 15 & $30.8 \pm 2.5\%$\\
30 & 80 & $47.3 \pm 2.5\%$\\
15 + 30 &  5 & $30.4 \pm 3.4\%$\\ 
15 + 30 & 80 & $60.8 \pm 1.2\%$\\
5 + 80 & 15 & $48.8 \pm 2.2\%$\\
5 + 80 & 30 & $59.4 \pm 3.2\%$\\
\bottomrule
\end{tabular}
}
\end{table}

\mybox{
The attack model can extract model information from t-SNE plots generated from different dataset distributions, density and perplexity settings.
}  

\subsection{Comparison with Existing Hyperparameter Stealing Attacks}
\label{subsection:comparison with existing model stealing attacks}

We compare our attack with the query-based model hyperparameter stealing attacks (referred to as query-based attacks)~\cite{OASF18}.
The experiment is conducted with the model type inference task on CIFAR-10.
The query-based attacks first train shadow models in the same way as our attack.
A set of 100 randomly selected images from the CIFAR-10 test set is then used to query the shadow models.
By querying each shadow model, the output posteriors for the 100 images are then concatenated, resulting in a 1,000-dimensional vector which is used as the input for the attack model.
The attack model is a multilayer perceptron (MLP) with 2 hidden layers of 1,000 neurons and trained with an SGD optimizer with a learning rate of 0.001 and momentum of 0.9 for 100 epochs.
\autoref{table:query attack performance} shows, as expected, the query-based attack performs very well on the model type inference task, reaching 99.7\% inference accuracy (ours is 92.8\%) when the adversary has access to complete posteriors information and with other hyperparameters fixed.

Note that the query-based attack needs to query the target model, which means the performance may be affected by limiting the output precision of the target model.
For instance, in the mixed setting (\autoref{table:query attack performance}), when the target model's response is the predicted label instead of the posteriors, the inference accuracy for the query-based attack is only 67.4\% while our attack can still achieve 77.1\% accuracy.

\begin{table}[!t]
\centering
\caption{Comparison with query-based attack.}
\label{table:query attack performance}
\scalebox{0.7}{
\begin{tabular}{c c c c} 
\toprule
\textbf{Attack Setting} & \textbf{Our Attack} & \textbf{Posteriors} &\textbf{Predicted Label}\\ 
\midrule
Fixed & $92.8 \pm 0.1$\% &  $99.7 \pm 0.1\%$ & $98.1 \pm 0.1\%$\\
Mixed & $77.1 \pm 1.0$\% &  $93.8 \pm 0.5\%$ & $67.4 \pm 0.8\%$\\
\bottomrule
\end{tabular}
}
\end{table}

\mybox{
Our attack is comparable to query-based attacks and even surpasses it when the target model's query output is limited.
}

\subsection{Downstream Adversarial Examples Attack}
\label{subsection:downstream attack}

We now demonstrate one of the potential use cases for our attack.
We consider an adversary who aims to cause the target model to misclassify data by crafting adversarial examples with only black-box and limited query access.
To achieve this goal, the adversary can craft adversarial examples from the surrogate model and transfer them to the target model.
We wonder whether the t-SNE plot generated from the target model can help the adversary build a good surrogate model.

For the following evaluation, we use the pre-trained model type setting in \autoref{subsection:inference targets for tsne attacks} on SVHN datasets.
1,300 images (10\% of the testing target dataset) are randomly sampled from the target testing dataset for crafting adversarial examples on the target models.
We first use our attack model to infer the model type, batch size, and optimization algorithm of the target model.
In this way, the adversary can minimize the interaction with the target model and keep the attack stealthy.
Then based on the inference result, we randomly select one of the shadow models that have matching hyperparameter settings and use it as the surrogate model to craft adversarial examples on the target model.
We assume that the adversary uses FGSM~\cite{GSS15} (a simple yet effective method) to alter a given image based on the gradient information from a given model.
The performance of this downstream attack is evaluated by comparing the misclassification rates (attack success rate) of the following 3 settings.
The white-box setting serves as a baseline, where the adversary has full knowledge of the target model.
The inferred model setting is our attack described above.
The random model setting is to mimic the adversary randomly selecting a shadow model and using it as the surrogate model.
\autoref{table:adversarial example} shows that using our inferred model, the adversary can craft better adversarial examples than those from a random shadow model.
For instance, the gap in attack success rate between inferred model setting and the white-box baseline is only around 7\%, while the gap between the random shadow model and the white-box baseline is around 10\%.
$\epsilon$ is the pixel-wise perturbation amount in FGSM that controls the strength of the noise added.
A higher $\epsilon$ leads to a higher attack success rate (see \autoref{table:adversarial example}), but the adversarial examples can be more apparent due to greater distortion.
The improvement of attack success rate when using our attack is consistent given all epsilon settings used in the experiments, which demonstrates the efficacy of our side-channel attack.

\mybox{
Our attack enables the attackers to generate adversarial examples more effectively by identifying a shadow model similar to the target model.
}

\begin{table}[!t]
\centering
\caption{Adversarial examples misclassification rate on target models crafted using different attack settings.}
\label{table:adversarial example}
\scalebox{0.7}{
\begin{tabular}{l c c c c} 
\toprule
\textbf{Attack Setting} & \textbf{$\epsilon=0.1$} & \textbf{$\epsilon=0.2$}& \textbf{$\epsilon=0.3$} \\ 
\midrule
White-box Target Model (Baseline) &  $61.6 \pm 4.9\%$ & $71.9 \pm 6.1\%$ & $77.4 \pm 6.4\%$  \\
Inferred Model (Our Attack) & $54.1 \pm 6.5\%$ & $64.8 \pm 6.0\%$ & $71.5 \pm 5.9\%$  \\
Random Shadow Model & $50.6 \pm 9.2\%$ & $61.3 \pm 7.8\%$ & $68.5 \pm 6.9\%$  \\
\bottomrule
\end{tabular}
}
\end{table} 

\subsection{t-SNE Defense}
\label{section:t-sne defense}

\begin{table}[!t]
\centering
\caption{Defense Overview. 
Note that we use A-B to denote a defense scenario, e.g., E-R denotes we conduct defense over \textbf{E}mbeddings via \textbf{R}ounding strategy.}
\label{table:defense_summary}
\scalebox{0.7}{
\begin{tabular}{l| c c } 
\toprule
\textbf{Strategy} & \textbf{Embeddings} & \textbf{Coordinates}\\ 
\midrule
\textbf{Rounding} & E-R (Security:\ding{55} Utility:\ding{51}) &  C-R (Security:\ding{51} Utility:\ding{55}) \\ 
\textbf{Thresholding} &  E-T (Security:\ding{51} Utility:\ding{51}) & NA \\
\textbf{Noising} &  E-N (Security:\ding{51} Utility:\ding{55}) &  C-N (Security:\ding{51} Utility:\ding{51})\\
\bottomrule
\end{tabular}
}
\end{table}

\begin{table}[!t]
\centering
\caption{Defense effectiveness against different strategies. 
The defenses are evaluated by the models trained on SVHN.
We highlight the successful defenses in \textbf{bold}.}
\label{table:defense}
\scalebox{0.7}{
\begin{tabular}{l c c c c} 
\toprule
\textbf{Defense Methods} & \textbf{Inference Acc.\ } & \textbf{Utility ($k$NN)} & \textbf{Utility (Visual)}\\  
\midrule
No Defense &  $74.1 \pm 1.0\%$ & 89.1\% & \ding{51}\\ 
E-R (0.1) & $70.2 \pm 1.8\%$ & 89.1\% & \ding{51}\\
E-R (INT) & $69.4 \pm 0.8\%$ & 89.1\% & \ding{51}\\
E-T (TOP 75\%) & $43.3 \pm 0.7\%$ & 88.6\% & \ding{51}\\
\textbf{E-T (TOP 60\%)} &  $33.9 \pm 1.3\%$ & 88.1\% & \ding{51}\\ 
E-N (1 x STD) & $36.6 \pm 1.2\%$ & 88.3\% & \ding{55}\\
E-N (0.5 x STD) & $55.2 \pm 2.3\%$ & 88.7\% & \ding{51}\\
C-R (to INT) & $67.2 \pm 1.6\%$ & 88.9\% & \ding{55}\\
C-R (to EVEN INT)& $41.0 \pm 3.1\%$ & 88.9\% & \ding{55}\\
C-N (2\% x STD) & $59.3 \pm 0.8\%$ & 89.1\% & \ding{51}\\
\textbf{C-N (5\% x STD)} & $37.8 \pm 2.1\%$ & 89.0\% & \ding{51}\\
\bottomrule
\end{tabular}
}
\end{table}

We consider the {\it disturbance-based approaches}, including rounding, thresholding, and noising, as the defense.
There are two main places to apply our defense strategies: introducing disturbances to {\it embedding values} used to fit the t-SNE, or directly to the {\it t-SNE coordinates}.
In total, we have 5 possible defense strategies as shown in \autoref{table:defense_summary} (thresholding coordinates is infeasible as the dimension of coordinates cannot be further reduced due to the visualization purpose).
We use the default experimental setting with SVHN as the dataset and model type as the inference target.
The defense is evaluated on security by the decrease in inference accuracy and on utility both quantitatively and visually.
Quantitatively, we use $k$NN accuracy as the utility metric~\cite{MH08}, where $k$NN accuracy is defined as the $k$-nearest neighbor classification accuracy with sample points’ coordinates in the given t-SNE as input and class labels as output.
If two t-SNE plots produce similar $k$NN accuracy, we can assume the 2 t-SNE plots represent the features in 2-dimensional space similarly.
We also evaluate the utility by visual observations, to determine if the protected t-SNE has noticeable deviations from the original version, which sometimes the $k$NN metric fails to capture.

\autoref{table:defense} shows that embedding thresholding (E-T) and coordinate noising (C-N) are two effective defense mechanisms as they largely reduce the attack success rate and preserve the utility.
For the t-SNE plots defended by C-N (5\% x STD), the inference accuracy decreases to 36.6\% while the $k$NN accuracy is extremely close to the original and the t-SNE with added noise shows almost no difference by visual examination (\autoref{figure:tsne defense} in Appendix).
We provide the detailed defense discussion below.

\mypara{Rounding Embedding Values (E-R)}
We conduct experiments on embedding with values rounded to specified decimal points, thus decreasing the resolution of embedding values.
We observe that E-R can preserve the utility but the attack accuracy remains relatively unperturbed (74.1\% to 70.2\%).

\mypara{Threshold Embedding Values (E-T)}
For this defense, in each embedding, only the largest $k$ percentage of values are maintained and the others are set to zero.
Then those modified embeddings are used to fit the t-SNE.
We find that, by setting a proper number of $k$, the defense can significantly reduce the attack performance while producing t-SNE plots that strongly resemble the original.
For example, when using only top 60\% embedding values, the inference performance decreases from 74.1\% to 33.9\% with less than a 1\% difference in $k$NN accuracy compared to the original.
Inspecting the t-SNE qualitatively also shows no noticeable difference from the original (see \autoref{figure:tsne defense} in Appendix).

\mypara{Gaussian Noise in Embedding (E-N)}
We also explore the effectiveness of adding Gaussian noise directly to embeddings.
Gaussian noise with standard deviation set to a percentage of the current embedding values’ standard deviation is added to the embedding before fitting the t-SNE.
The defense is unsuccessful even with a larger standard deviation.
With the added noise having 50\% of the embedding values’ standard deviation, the inference performance decreases to 55.2\%, while maintaining a $k$NN accuracy same as the original.

\mypara{Rounding t-SNE Point Coordinates (C-R)}
Instead of rounding embedding values used to fit t-SNE, we directly round the sample point’s coordinates and thus diminish clusters' geometric characteristics.
\autoref{table:defense} shows promising $k$NN classification accuracy for both rounding to integers and rounding to even integers.
Rounding to an even integer also greatly mitigates attack effectiveness.
However, \autoref{figure:tsne defense} shows the rounding effect can be easily detected from observation.
t-SNE with rounding shows a distinct grid pattern when the rounding unit is high (low rounding unit does not provide security).
For example, even for rounding to the integer, the grid pattern is already noticeable.

\mypara{Gaussian Noise in t-SNE Point Coordinates (C-N)}
Similar to adding noise to embedding values, we add Gaussian noise directly to t-SNE points’ coordinates to disturb the distinct patterns in cluster shapes.
The standard deviation is set to a percentage of the overall coordinates’ standard deviation in the current plot.
\autoref{table:defense} shows this method can effectively mitigate model information extraction from t-SNE plots.
With Gaussian noise of 5\% original standard deviation, the inference performance decreases to 37.8\%.
This method also produces t-SNE that strongly resembles the original version.
$k$NN accuracy is extremely close to the original and the t-SNE with added noise shows almost no difference by visual examination (see \autoref{figure:tsne defense} in Appendix).

\mybox{
Thresholding embedding values and adding Gaussian noise in the t-SNE point coordinates are the two most effective defenses against the attack on t-SNE plots.
}

\subsection{Adaptive Attack}
\label{section:Adaptive Attack}

We then consider an adaptive attacker who is aware of the effective defense mechanisms used in the t-SNE plots.
In this way, they can train the attack model on the original t-SNE plots and those t-SNE plots altered by the effective defense methods, i.e., top 60\% embedding values thresholding and coordinates noising (5\% STD).
The adaptive attack can successfully render both defense methods ineffective, as seen in \autoref{table:adaptive tsne}.
The inference accuracy of the adaptive attack model improves drastically on both protected data, achieving similar performance as unprotected data.
The adaptive attack also performs well on other less effective defense methods that use the same strategy without retraining.
For example, the inference accuracy for top 75\% embedding thresholding and 2\% STD coordinate noising are both around 66\%, which is a 18\% and 8\% increase respectively.
The adaptive attack further demonstrates that the privacy risk of model information leakage from t-SNE plots is underestimated.

\begin{table}[!t]
\centering
\caption{Adaptive t-SNE attack performance on SVHN.}
\label{table:adaptive tsne}
\scalebox{0.7}{
\begin{tabular}{l c c c} 
\toprule
\textbf{Defense Methods} & \textbf{Inference Acc.\ } & \textbf{Inference Acc.\ (non-adaptive)}\\  
\midrule
No Defense &  $72.5 \pm 1.9\%$ & $74.1 \pm 1.0\%$ \\ 
\textbf{E-T (TOP 60\%)} &  $83.4 \pm 0.8\%$ & $36.5 \pm 2.4\%$ \\ 
E-T (TOP 75\%) &  $65.6 \pm 1.1\%$ & $47.1 \pm 1.2\%$ \\ 
\textbf{C-N (5\% x STD)} & $66.7 \pm 1.3\%$ & $41.1 \pm 2.7\%$ \\
C-N (2\% x STD) & $67.7 \pm 1.2\%$ & $59.1 \pm 1.2\%$ \\
\bottomrule
\end{tabular}
}
\end{table}

\mybox{
With adaptive attacks, our model nullifies two most effective defense methods proposed.
}

\section{Loss Plot Evaluation}
\label{section:loss plot}

In this section, we demonstrate the attack is not limited to t-SNE plots but is also capable of attacking loss plots, which is another type of scientific plot widely used to showcase model convergence performance over the training process.
  
\subsection{Attack Performance}
\label{subsection:loss plot model type}

The attack follows the same attack methodology and uses the same default settings for shadow, target, and attack model training as previous t-SNE attacks.
The inference targets are also similar, including model type (6 pre-trained), batch size, and optimization algorithms on CIFAR-10, SVHN, and FashionMNIST datasets.
We generate loss plots both with and without axis information using settings from \autoref{section:evaluation setup} (see \autoref{figure:loss plot defense} for an example).
\autoref{table:loss plot inference performance} shows that our attack can successfully infer model information from loss plots as well.
The attack accuracy on the three types of inference targets are generally better than those on t-SNE plots.
The results are expected since a loss plot is a direct reflection of the model's behavior.
We also observe that additional axis information improves attack performance.
The average model type inference accuracy on loss plots with axis information generated from models trained on SVHN is 86.8\%.
For the loss plots without axis information, our attack can still achieve 76.2\% accuracy, which shows that the loss curve itself plays an important role in the attack model to distinguish different model types.
The confusion matrix of the inference results is shown in \autoref{figure:confusion plot loss} (in Appendix).

\begin{table}[!t]
\centering
\caption{The average attack accuracy on loss plots for different inference targets on 3 different datasets. 
The values in the parenthesis are results for loss plots without axis.}
\label{table:loss plot inference performance}
\tabcolsep 3pt
\scalebox{0.7}{
\begin{tabular}{l c c c c} 
\toprule
\textbf{Inference Targets} & \textbf{CIFAR-10} & \textbf{SVHN} & \textbf{FashionMNIST}  \\ 
\midrule
Model Type <6> & $87.1(78.4) \pm 0.9\%$ & $86.8(76.2) \pm 0.8\%$ & $71.9(56.7) \pm 0.3\%$ \\
Batch Size <4> & $91.4(90.5) \pm 0.4\%$ & $90.2(90.1) \pm 0.5\%$ & $86.9(86.2) \pm 0.6\%$ \\
Optimizer <2>  & $98.4(96.7) \pm 0.2\%$ & $99.7(98.5) \pm 0.4\%$ & $84.0(74.6) \pm 0.3\%$ \\
\bottomrule
\end{tabular}
}
\end{table} 

\subsection{Loss Plot Defense}
\label{subsection:loss plot defense}

To prevent the model information leakage from loss plots, we consider three defenses below, i.e., Gaussian noise, TensorBoard smoothing, and sliding window smoothing.
We use the $L_2$ distance between original loss curves and the protected ones as the utility of the defense (a successful defense should not destroy the usefulness of the original plot, and thus should have a high utility).

\begin{table}[!t]
\centering
\caption{Loss plot defense performance on SVHN.}
\label{table:loss defense}
\scalebox{0.7}{
\begin{tabular}{l c c c} 
\toprule
\textbf{Defense Methods} & \textbf{Acc.\ w Axis} & \textbf{Acc.\ wo Axis} & \textbf{$L_2$ Distance}\\ 
\midrule
No Defense &  $86.8 \pm 0.8\%$ & $76.2 \pm 0.7\%$ & 0\\ 
Gaussian Noise & $48.5 \pm 1.9\%$ & $48.4 \pm 2.1\%$ & 1.211\\
TensorBoard & $81.1 \pm 0.9\%$ & $65.2 \pm 1.1\%$ & 0.310\\
Sliding Window & $44.5 \pm 2.6\%$ & $57.1 \pm 1.8\%$ & 0.827\\
\bottomrule
\end{tabular}
}
\end{table}

\begin{figure}[!t]
\centering
\begin{subfigure}{0.118\textwidth}
\centering
\includegraphics[width=\textwidth]{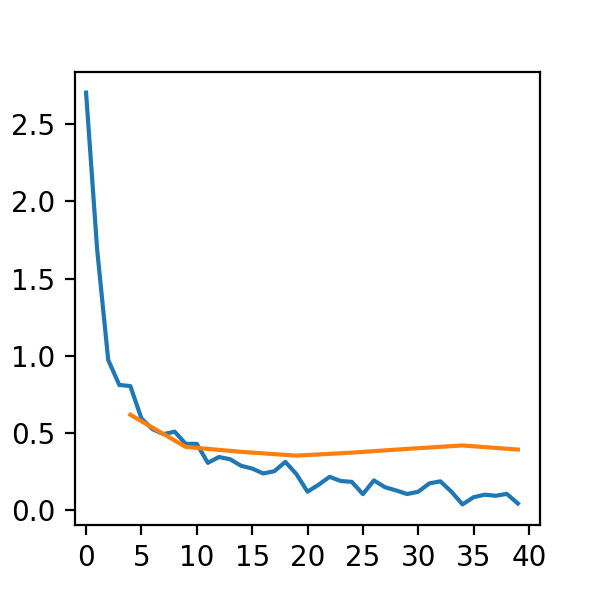}
\end{subfigure}%
\begin{subfigure}{0.118\textwidth}
\centering
\includegraphics[width=\textwidth]{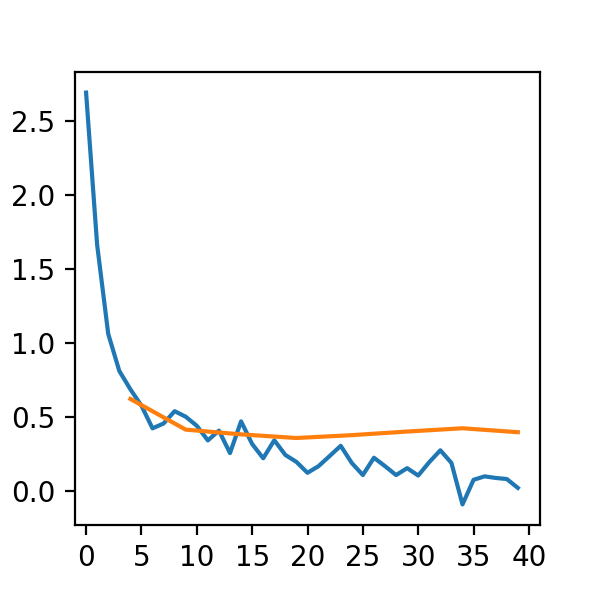}
\end{subfigure}%
\centering
\begin{subfigure}{0.118\textwidth}
\centering
\includegraphics[width=\textwidth]{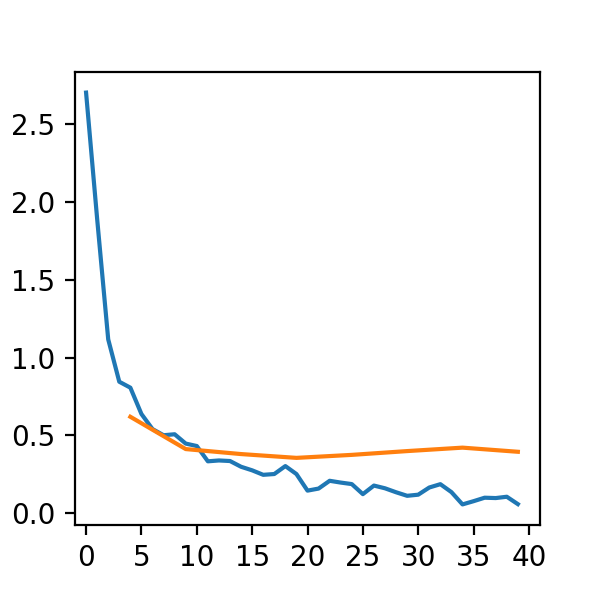}
\end{subfigure}%
\begin{subfigure}{0.118\textwidth}
\centering
\includegraphics[width=\textwidth]{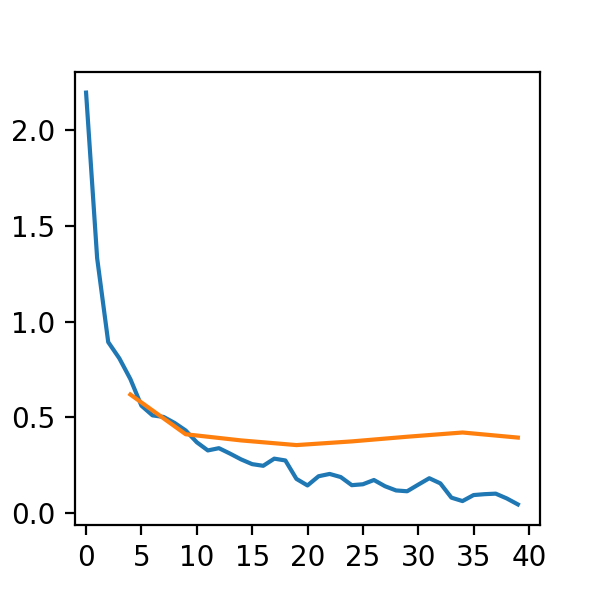}
\end{subfigure}%

\centering
\begin{subfigure}{0.118\textwidth}
\centering
\includegraphics[width=\textwidth]{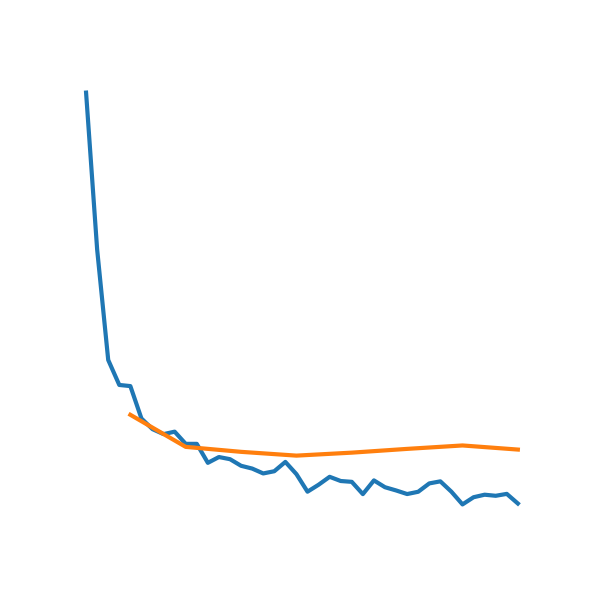}
\caption{O}
\label{figure:svhn defense og}
\end{subfigure}%
\begin{subfigure}{0.118\textwidth}
\centering
\includegraphics[width=\textwidth]{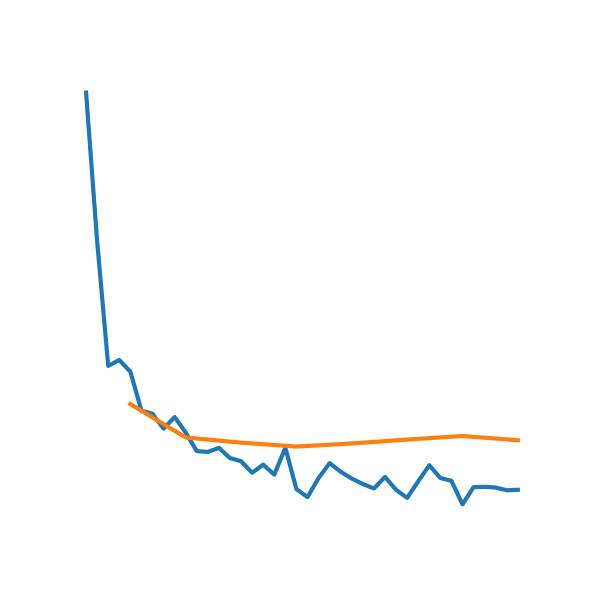}
\caption{N}
\label{figure:svhn defense gaussian}
\end{subfigure}%
\centering
\begin{subfigure}{0.118\textwidth}
\centering
\includegraphics[width=\textwidth]{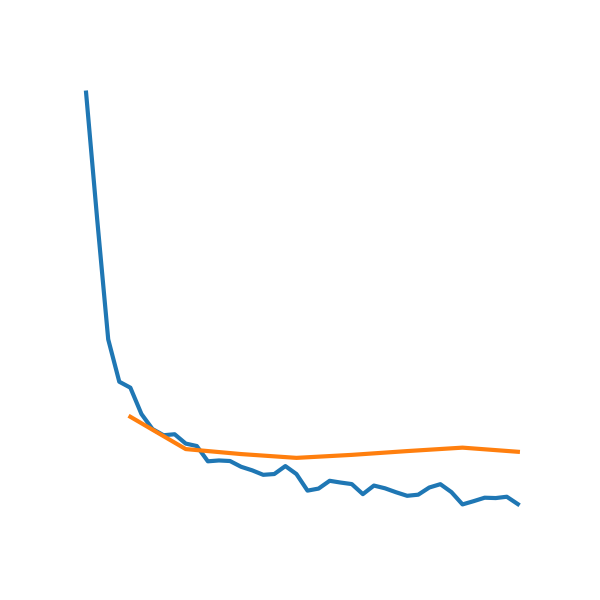}
\caption{T}
\label{figure:svhn defense tensor}
\end{subfigure}%
\begin{subfigure}{0.118\textwidth}
\centering
\includegraphics[width=\textwidth]{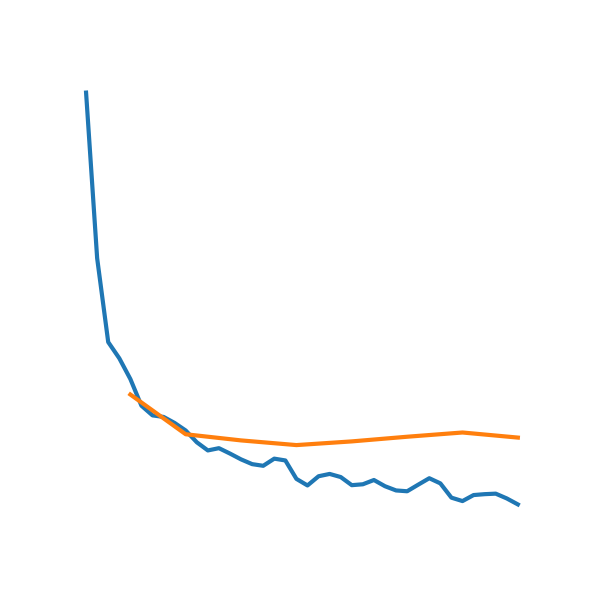}
\caption{S}
\label{figure:svhn defense slide}
\end{subfigure}%
\caption{Loss plots with different defenses.
Here O denotes the \textbf{O}riginal model without any defense, N denotes the Gaussian \textbf{N}oise, T denotes \textbf{T}ensorboard Smoothing, and S denotes \textbf{S}liding Window Smoothing.
For the loss plots with axis information, the x-axis denotes the timestamp (an epoch is divided into 5 timestamps) and the y-axis denotes the loss value.}
\label{figure:loss plot defense}
\end{figure}

\mypara{Gaussian Noise}
One way to introduce disturbance in the loss curve while preserving overall characteristics is to add Gaussian noise to the losses.
Here we use the average standard deviation of loss values as the standard deviation of the Gaussian noise.
As we can see in \autoref{table:loss defense}, the results show that adding Gaussian noise can mitigate the inference performance.
The inference accuracy is reduced by around 40\% on loss plots with the axis.
The mitigation is less effective on loss plots without the axis.
The $L_2$ distance between the original and altered loss curve is 1.211, the highest among all 3 defense methods.
When inspecting the loss curve visually in \autoref{figure:loss plot defense}, we observe the defense has altered crucial information in the loss curve.
For instance, at around $15$-th timestamp, the training loss is notably higher than the test loss, which is not a characteristic of the original loss curve.
Adding Gaussian noise is thereof not ideal.
  
\mypara{TensorBoard Smoothing}
We apply the loss curve smoothing strategy used in TensorBoard~\cite{TensorBoard}, a popular tool among machine learning researchers.
The loss value at each timestamp is averaged between the loss value at the current and previous timestamps, with a scalar constant controlling the weight of each value: $\mathcal{L}^*_t = w\mathcal{L}_{t-1} + (1-w) \mathcal{L}_t$,
where $w$ is a weight factor.
With the weight factor set at 0.2, i.e., with 80\% of the weight on the current value, the defense is not effective, decreasing the inference performance by only 5.7\% and 11.0\% respectively with and without axis in loss plots, although both observations and the quantitative $L_2$ distance show TensorBoard smoothing does have high utility.
  
\mypara{Sliding Window Smoothing}
Another smoothing technique is using a sliding window.
The smoothed value at a given timestamp $t$ is calculated by averaging the loss value starting from timestamp $t$ till timestamp $t+s$, where $s$ is the sliding window size.
If the endpoint is beyond total timestamps, the average is calculated with existing values: $\mathcal{L}^*_t = \dfrac{1}{n}\sum_{n=t}^{t+s} \mathcal{L}_n$.
With a window size of 2, sliding window smoothing provides good protection against model information extraction from loss plots, compared to the previous two methods.
The inference accuracy decreases for both types of loss plots.
For instance, the attack accuracy decreases 42.3\% on loss plots with axis.
Although the $L_2$ distance is higher than that of Tensorboard smoothing, the loss curve's overall characteristics are largely preserved (by observing \autoref{figure:svhn defense slide}).
The altered loss curve still presents the model's convergence rate, overfitting level, and general performance accurately.

\subsection{Adaptive Attack}

\begin{table}[!t]
\centering
\caption{Adaptive loss plot attack performance.
The value in the parenthesis denotes the original attack performance.}
\label{table:adaptive loss attack}
\scalebox{0.7}{
\begin{tabular}{l c c} 
\toprule
\textbf{Defense Methods} & \textbf{Acc.\ w Axis} & \textbf{Acc.\ wo Axis}\\ 
\midrule
No Defense &  $89.5(86.8) \pm 0.5\%$ & $77.5(76.2) \pm 0.9\%$ \\ 
TensorBoard & $89.6(81.1) \pm 0.4\%$ & $67.7(65.2) \pm 0.9\%$ \\
Sliding Window & $84.1(44.5) \pm 0.7\%$ & $59.2(57.1) \pm 2.1\%$\\
\bottomrule
\end{tabular}
}
\end{table}

Similar to the adaptive attack on t-SNE (see \autoref{section:Adaptive Attack}), we assume the adaptive adversary has knowledge of the defense methods deployed.
The adaptive attack model is trained on the original loss plots and the ones with the two defense methods, sliding window smoothing and TensorBoard smoothing (although TensorBoard smoothing is not effective, for completeness, we assume the defense might be selected for its high utility).
The attack performance is evaluated on the original testing loss plots and the two altered versions respectively.
\autoref{table:adaptive loss attack} shows the adaptive attack achieves high accuracy across all settings.
For instance, the inference accuracy on loss plots with the axis reaches 84.1\% on a dataset protected with sliding window smoothing.
It represents an increase of 39.6\% compared to the original attack.
The adaptive attack also improves inference accuracy on loss plots with axis more than those without.
This is because the defense methods are not as effective on the loss plots without axis in the first place.
Interestingly, the adaptive model performs better on the no-defense dataset as well.
We suspect the altered loss plots serve as data augmentation and improve the attack model's generalization ability.
The high accuracy of the adaptive attack further accentuates the potential threat of model information stealing attacks from loss plots.

\section{Grad-CAM Analysis}
\label{section:grad-cam analysis}

\textbf{Grad}ient-weighted \textbf{C}lass \textbf{A}ctivation \textbf{M}apping (Grad-CAM) is a popular technique that provides visual explanations for CNNs~\cite{SCDVPB17}.
We use Grad-CAM on the attack models to analyze features in scientific plots that enable successful model information inference.

\mypara{t-SNE Plots Analysis}
When conducting the ablation study on t-SNE (see \autoref{subsection:ablation study}), we notice that the attack models’ performance remains similar across different color settings.
We deduce that the attack model extracts information from t-SNE mostly from the clusters’ geometric shapes.
\autoref{figure:gradcam tsne} (in Appendix) shows Grad-CAM heatmaps of selected samples in model type inference attack.
We observe that the attack model can identify unique characteristics in t-SNE plots for inference targets based on the heatmaps.
While the attack model uses the scattered clusters to classify MobileV3's t-SNE, it finds more distinct characteristics among similar model types, such as those from the ResNet family.
Grad-CAM visualization also explains the high misclassification rate between ResNet34 and ResNet50 (as seen in \autoref{figure:confusion cifar10} in Appendix).
These two model types are very similar and the patterns identified by the attack model from the t-SNE plots (see \autoref{figure:gradcam tsne resnet34} and \autoref{figure:gradcam tsne resnet50}) are also hard to distinguish by human eyes.
It is therefore understandable that such slight variations can indeed cause confusion between these two geometric features.

\mypara{Loss Plots Analysis}
\autoref{figure:gradcam loss without axis} and \autoref{figure:gradcam loss with axis} (in Appendix) show Grad-CAM heatmaps on loss plot attacks.
For both with and without axis loss plots, the attack model utilizes mostly the loss information in the first 10 timestamps.
When the loss plot includes axis, the attack model uses the added quantitative information to improve classification performance for certain classes, e.g., ResNet34 or ResNet50 (the two highest misclassified model types without axis, see \autoref{figure:confusion plot loss} in Appendix).
Grad-CAM visualization shows that the attack model can correctly identify regions in loss plots most relevant to the original model's performance.

\section{Related Work}
\label{section:related_work}

Previous research has shown that machine learning models are vulnerable to model stealing attacks~\cite{TZJRR16,JCBKP20,CJM20,WG18,OASF18,OSF19,KTPPI20,SHHZ22,SHYBZ22}.
The core assumption of those attacks is that the adversary has black-box access to the target model and then launches stealing attacks via the query-response information.
They mainly focus on extracting the target model's parameters~\cite{CJM20,TZJRR16,JCBKP20}, hyperparameters~\cite{WG18,OASF18}, and functionalities~\cite{OSF19,KTPPI20,SHHZ22,JCBKP20,SHYBZ22}.
Such attacks have also been applied to different machine learning paradigms such as NLP~\cite{KTPPI20}, Graph Neural Networks (GNNs)~\cite{SHHZ22}, and Contrastive Learning~\cite{SHYBZ22}.
The closest work to ours is Shen et al.~\cite{SHHZ22}.
It shows that a query-based attack can be conducted to steal a target model's (GNN) functionality when the model provides the t-SNE coordinates as the response.
Different from Shen et al.~\cite{SHHZ22}, our attack aims at stealing model information instead of the model itself.
More importantly, we show that even one single scientific plot is enough to reveal the hyperparameters of the target model.

To mitigate the model stealing attacks, several defense mechanisms have been proposed~\cite{JSMA19,OSF20,LEMS19,KQ20,JCCP21,CHZ22}.
Broadly speaking, existing defenses focus on query-based model stealing attacks and can be classified into two categories.
The first category centers on reactively analyzing the querying data.
These approaches prevent model stealing attacks by raising alerts when the query data's distribution largely deviates from the overall benign query data distribution~\cite{JSMA19}, giving incorrect predictions to OOD query samples~\cite{KQ20}, or embedding watermarks into the target model so that a model owner can later prove ownership~\cite{JCCP21,CHZ22}, etc.
The second category aims at proactively defending the model stealing attack by using output perturbation~\cite{OSF20}, differential privacy~\cite{ZYHFS19}, model refinement~\cite{LEMS19}, etc.
We reveal that output perturbation techniques can effectively reduce our attack accuracy.
However, such defenses can still be bypassed by adaptive attacks.

\section{Limitations and Discussion}
\label{section:limiation}

We acknowledge that our work has some limitations.
First of all, most of our evaluations assume the attacker has the knowledge of the target model's training data distribution.
When directly evaluated on out-of-distribution data, the attack does not generalize well.
However, domain shift is one of the biggest challenges when deploying machine learning models in the real world \cite{HG17}.
This certainly applies to our attack models as well.
Also, our attack follows previous work's threat model where the training data distribution is not private~\cite{WG18,OASF18}.
We further show that our attack is still potent when fine-tuned using a small amount of in-distribution data.

Secondly, in the pre-trained model architecture settings, the evaluation is conducted on 6 model types belonging to 3 families.
Though these models are the most popular ones, there are many more model types and families that we have yet to examine our attack on.
However, the high attack accuracy in predicting model families indicates that it is easier for our attack to distinguish models from different architecture families.
Models belonging to different families typically generate more distinct scientific plots, and thus, our attack should scale well with models from more architecture families.

Thirdly, all our experiments are conducted on CNN models.
We also perform our attacks on another type of machine learning model, namely, graph neural networks (GNNs).
For the model type inference, we achieve a 95.4\% accuracy, the concrete result is listed in \autoref{section:gnn_results}.
We plan to explore the effectiveness of our attacks against other types of machine learning models more thoroughly in the future.

\section{Conclusion}
\label{section:conclusion}

In this paper, we perform the first model information stealing attack against CNN models through scientific plots.
Empirical evaluation shows that our attack is effective in inferring the configurations of target models.
Our results also indicate that some defenses can effectively mitigate the attack.
However, those defenses fail when an adaptive attacker is considered.
This further demonstrates the severe risk of scientific plots leaking target model's information.
We hope our discovery of model information leakage from scientific plots can inspire future work to develop more robust ones against such attacks.

\medskip
\mypara{Acknowledgments}
We thank all anonymous reviewers for their constructive comments. 
This work is partially funded by the Helmholtz Association within the project ``Trustworthy Federated Data Analytics'' (TFDA) (funding number ZT-I-OO1 4), by the European Health and Digital Executive Agency (HADEA) within the project ``Understanding the individual host response against Hepatitis D Virus to develop a personalized approach for the management of hepatitis D'' (D-Solve) (grant agreement number 101057917), and by NSF grant number 2217071.

\begin{small}
\bibliographystyle{plain}
\bibliography{normal_generated_py3}
\end{small}

\appendix
\section*{Appendix}

\section{Procedure of t-SNE}
\label{section:introduction2tsne}

t-SNE has two main phases.
First, given the data's high-dimensional embeddings $\{{h_i}\}$, we define the similarity between each pair of embeddings as $p_{ij} = \frac{p_{i|j} + p_{j|i}}{2n}$ where 
\begin{align*}
p_{j|i} = \frac{\exp(- \| h_i - h_j \|^2/2 \sigma_i^2)}{\sum\limits_{k\neq i} \exp(- \| h_i - h_k \|^2/2 \sigma_i^2)}  
\end{align*}
and $\sigma_i$ is determined by the perplexity of the similarities.

Second, t-SNE finds the low-dimensional points $\{{\ell_i}\}$ corresponding to $\{{h_i}\}$ that minimizes the KL-divergence $\sum_i \sum_j p_{ij} \log \frac{p_{ij}}{q_{ij}}$, where $q=\{q_{ij}\}$ is the similarities among $\{{\ell_i}\}$, and is defined as
\begin{align*}
q_{ij} = \frac{(1 + \| \ell_i - \ell_j \|^2)^{-1}}{\sum\limits_{ s \neq t} (1 + \| \ell_s - \ell_t \|^2)^{-1}}    
\end{align*}
$\{{\ell_i}\}$ should have the same similarity characteristic (i.e., pairs close in high-dimensional space would also be close in low-dimensional space) as $\{{h_i}\}$.

\section{GNN Experiment Results}
\label{section:gnn_results}

While our work focuses on conducting attacks against CNN-based image classifiers, we believe the vulnerability exists in scientific plots for other types of data and models as well.
We conduct an experiment predicting the model types (GraphSAGE~\cite{HYL17}, GAT~\cite{VCCRLB18}, GIN~\cite{XHLJ19}) of graph neural network  (GNN) models trained on CiteSeer-Full~\cite{GBL98} using their corresponding t-SNE plots.
Our attack achieves 95.4\% accuracy with 30 shadow models.
The attack model transferred from previous experiments (CNN models) achieves 87.1\% accuracy even with only 3 shadow models (one for each model type).
The high transferability indicates the attack model can learn features from t-SNE plots that are useful for extracting target models' information even when the original target models are extremely different.

\section{Additional Figures}
\label{section:Addition Figures}

Here we include additional figures related to our study.

\begin{figure}[hbtp]
\centering
\begin{subfigure}{0.125\textwidth}
\centering
\includegraphics[width=\textwidth]{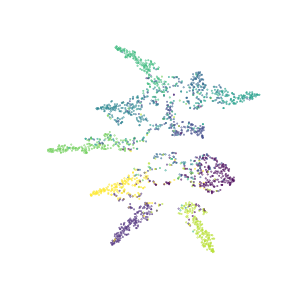}
\caption{Adam}
\label{figure:tsne adam}
\end{subfigure}%
\begin{subfigure}{0.125\textwidth}
\centering
\includegraphics[width=\textwidth]{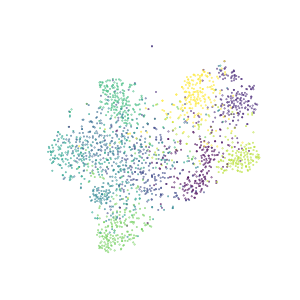}
\caption{SGD}
\label{figure:tsne sgd}
\end{subfigure}%
\caption{t-SNE plots of ResNet18 models trained with different optimization algorithms on CIFAR-10.}
\label{figure:optimization tsne}
\end{figure}

\begin{figure}[hbtp]
\centering
\begin{subfigure}{0.125\textwidth}
\centering
\includegraphics[width=\textwidth]{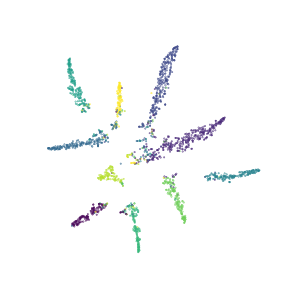}
\caption{Color}
\label{figure:color}
\end{subfigure}%
\begin{subfigure}{0.125\textwidth}
\centering
\includegraphics[width=\textwidth]{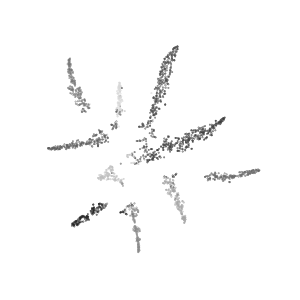}
\caption{Grayscale}
\label{figure:grayscale}
\end{subfigure}%
\begin{subfigure}{0.125\textwidth}
\centering
\includegraphics[width=\textwidth]{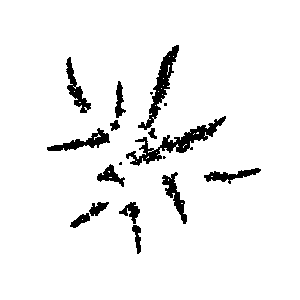}
\caption{Binary}
\label{figure:binary}
\end{subfigure}
\caption{t-SNE plots with different color settings. The model is ResNet18 trained on SVHN.}
\label{figure:tsne color}
\end{figure}

\begin{figure}[hbtp]
\centering
\begin{subfigure}{0.125\textwidth}
\centering
\includegraphics[width=\textwidth]{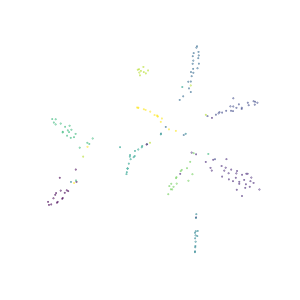}
\caption{200}
\label{figure:density 200}
\end{subfigure}%
\begin{subfigure}{0.125\textwidth}
\centering
\includegraphics[width=\textwidth]{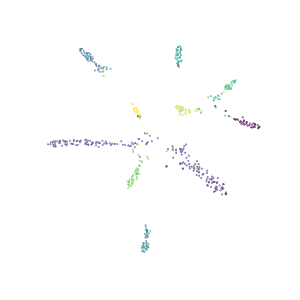}
\caption{500}
\label{figure:density 500}
\end{subfigure}%
\begin{subfigure}{0.125\textwidth}
\centering
\includegraphics[width=\textwidth]{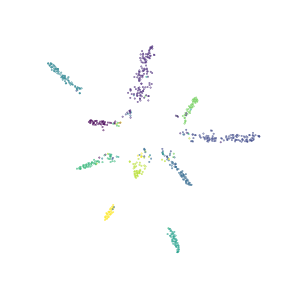}
\caption{700}
\label{figure:density 700}
\end{subfigure}

\begin{subfigure}{0.125\textwidth}
\centering
\includegraphics[width=\textwidth]{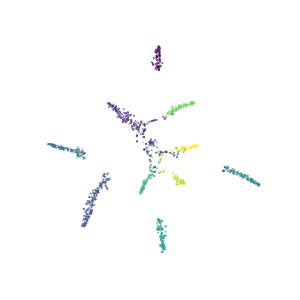}
\caption{1000}
\label{figure:density =1000}
\end{subfigure}%
\begin{subfigure}{0.125\textwidth}
\centering
\includegraphics[width=\textwidth]{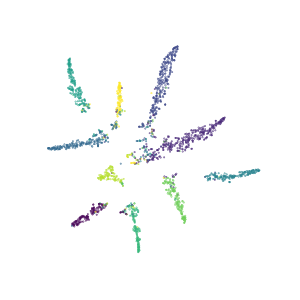}
\caption{2000}
\label{figure:density 2000}
\end{subfigure}%
\begin{subfigure}{0.125\textwidth}
\centering
\includegraphics[width=\textwidth]{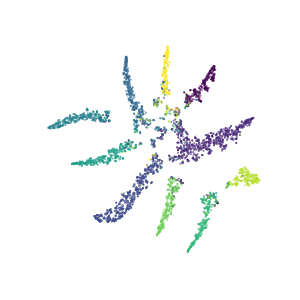}
\caption{4000}
\label{figure:density 4000}
\end{subfigure}
\caption{t-SNE plots with different sample densities.
The model is ResNet18 trained on SVHN.}
\label{figure:tsne density}
\end{figure}

\begin{figure}[hbtp]
\centering
\begin{subfigure}{0.125\textwidth}
\centering
\includegraphics[width=\textwidth]{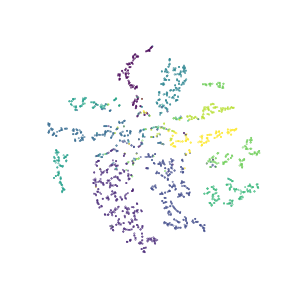}
\caption{$ 5$}
\label{figure:resnet50_perp5}
\end{subfigure}%
\begin{subfigure}{0.125\textwidth}
\centering
\includegraphics[width=\textwidth]{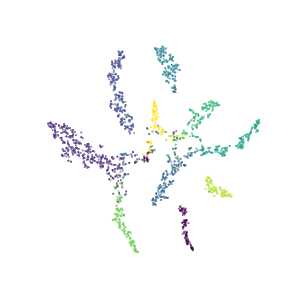}
\caption{$ 15$}
\label{figure:resnet50_perp15}
\end{subfigure}%
\centering
\begin{subfigure}{0.125\textwidth}
\centering
\includegraphics[width=\textwidth]{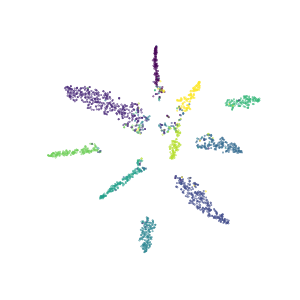}
\caption{$ 30$}
\label{figure:resnet50_perp30}
\end{subfigure}%
\begin{subfigure}{0.125\textwidth}
\centering
\includegraphics[width=\textwidth]{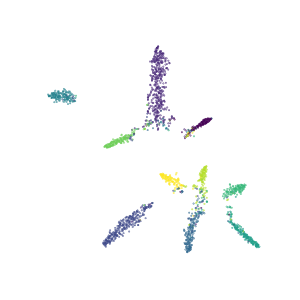}
\caption{$ 80$}
\label{figure:resnet50_perp80}
\end{subfigure}%
\caption{t-SNE plots with different perplexity values.
The model is ResNet50 trained on SVHN.}
\label{figure:tsne perplexity}
\end{figure}

\begin{figure}[hbtp]
\centering
\begin{subfigure}{0.16\textwidth}
\centering
\includegraphics[width=0.8\textwidth]{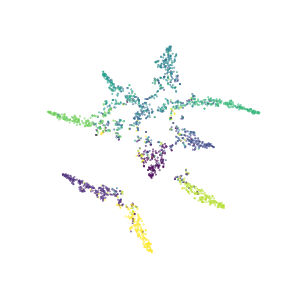}
\caption{CIFAR10}
\label{figure:resnet18_cifar10}
\end{subfigure}%
\begin{subfigure}{0.16\textwidth}
\centering
\includegraphics[width=0.8\textwidth]{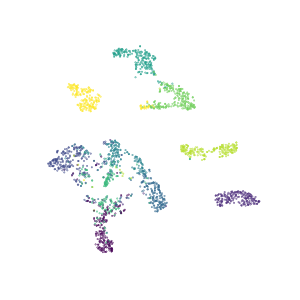}
\caption{FashionMNIST}
\label{figure:resnet18_fashion}
\end{subfigure}%
\begin{subfigure}{0.16\textwidth}
\centering
\includegraphics[width=0.8\textwidth]{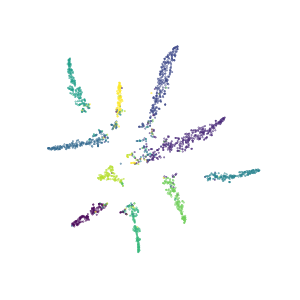}
\caption{SVHN}
\label{figure:resnet18_svhn}
\end{subfigure}
\caption{t-SNE plots of ResNet18 trained on different datasets.}
\label{figure:tsne datasets}
\end{figure}

\begin{figure}[hbtp]
\centering
\includegraphics[width=0.75\columnwidth]{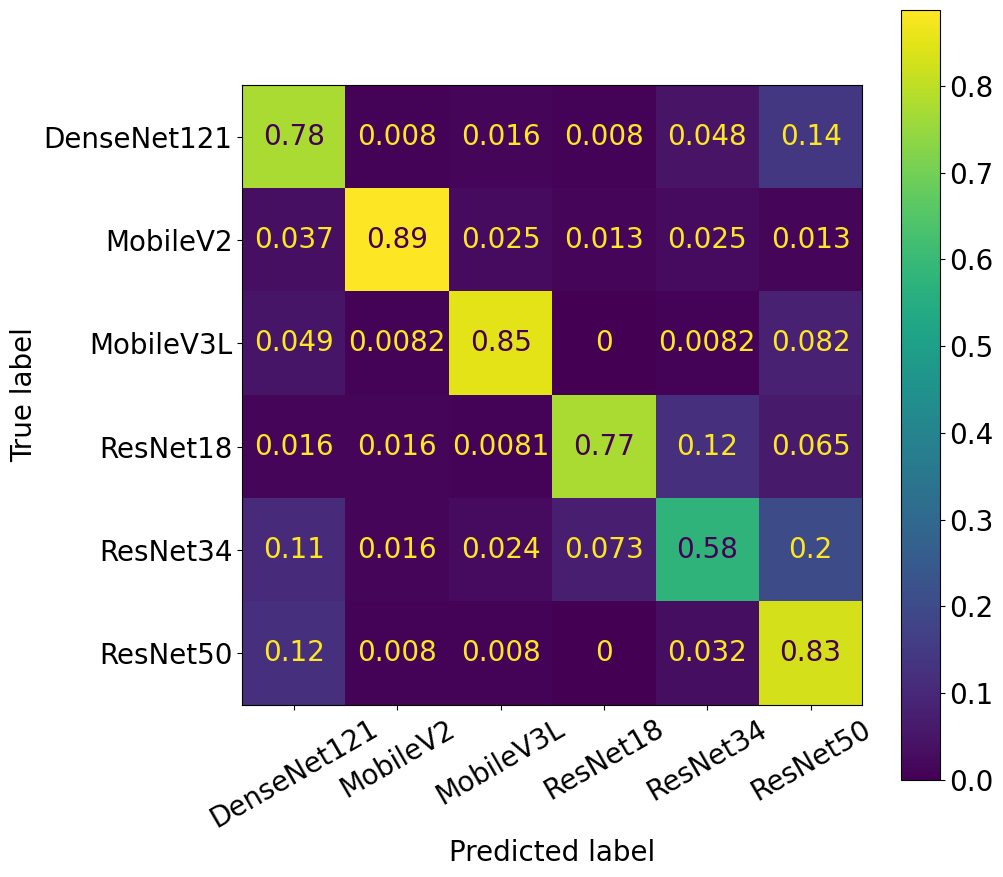}  
\caption{Confusion matrix for model type inference of t-SNE plots on CIFAR-10.
The inference accuracy is at least 58\% on each model type. 
The highest confusion occurs between ResNet34 and ResNet50.}
\label{figure:confusion cifar10}
\end{figure}

\begin{figure}[hbtp]
\centering
\includegraphics[width=0.75\columnwidth]{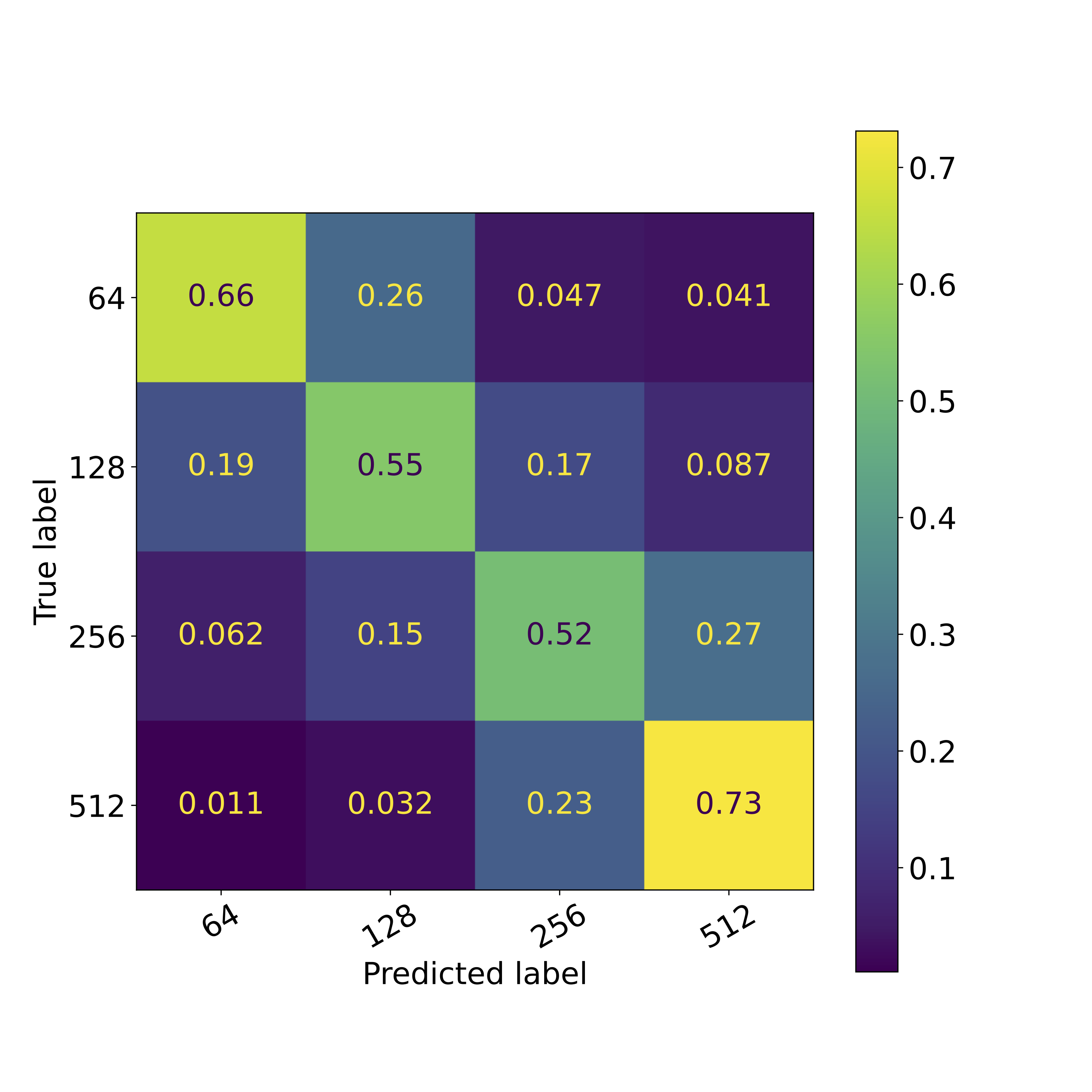}  
\caption{Confusion matrix for batch size inference of t-SNE plots on CIFAR-10.
Similar batch sizes are more easily confused.}
\label{figure:confusion bs}
\end{figure}

\begin{figure*}[hbtp]
\centering
\begin{subfigure}{0.75\columnwidth}
\centering
\includegraphics[width=\textwidth]{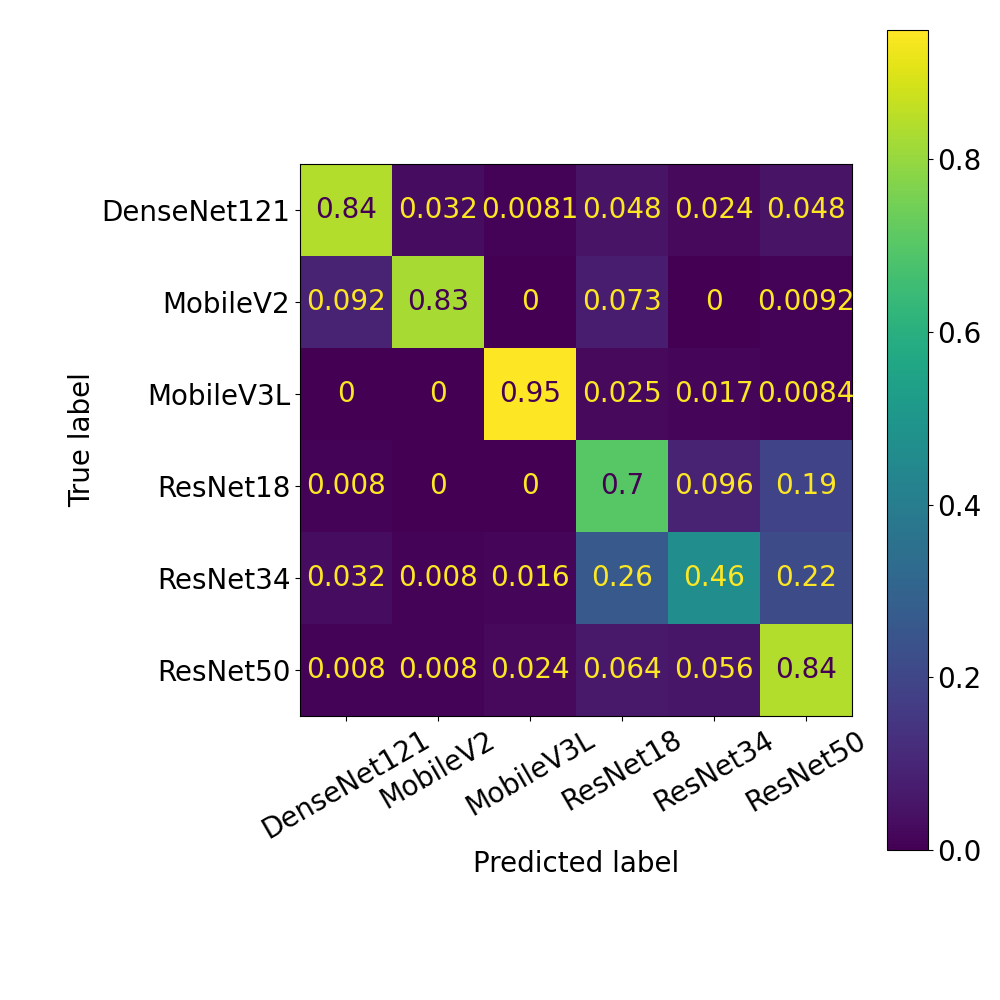}
\caption{Confusion Maxtrix for Loss Plots w/o Axis}
\end{subfigure}
\hfill
\begin{subfigure}{0.75\columnwidth}
\centering
\includegraphics[width=\textwidth]{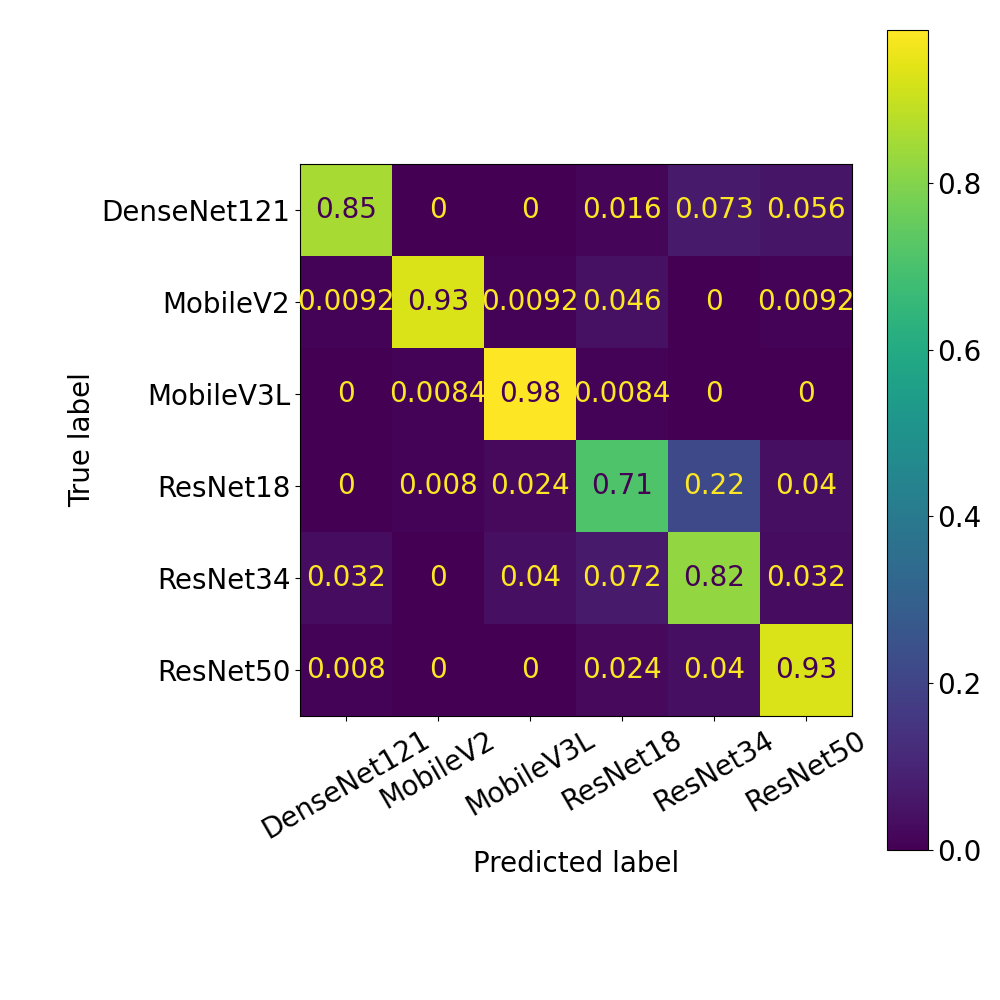}
\caption{Confusion Maxtrix for Loss Plots with Axis}
\end{subfigure}
\caption{Adding axis information reduces misclassification within the ResNet family models, especially for ResNet34 and ResNet50.}
\label{figure:confusion plot loss}
\end{figure*}

\begin{figure*}[t]
\centering
\begin{subfigure}[t]{0.24\textwidth}
\centering
\includegraphics[width=0.52\textwidth]{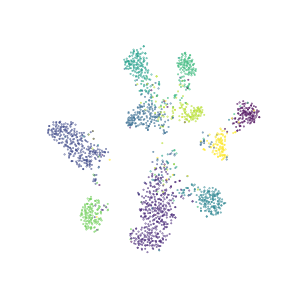}
\caption{Embedding Rounding}
\label{figure:embedround}
\end{subfigure}
\hfill
\begin{subfigure}[t]{0.24\textwidth}
\centering
\includegraphics[width=0.52\textwidth]{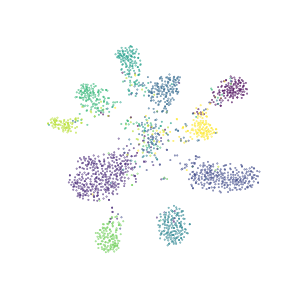}
\caption{Top 75\% Embedding}
\label{figure:top75 embed}
\end{subfigure}
\hfill
\begin{subfigure}[t]{0.24\textwidth}
\centering
\includegraphics[width=0.52\textwidth]{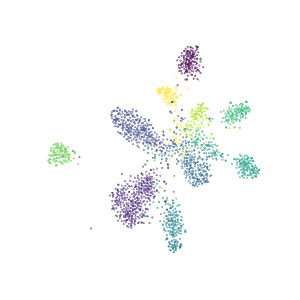}
\caption{Top 60\% Embedding}
\label{figure:top60 embed}
\end{subfigure}
\hfill
\begin{subfigure}[t]{0.24 \textwidth}
\centering
\includegraphics[width=0.52\textwidth]{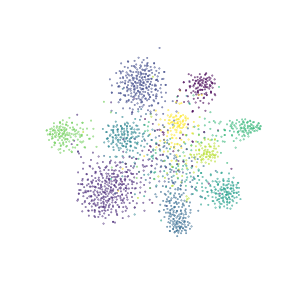}
\caption{Embedding Noise}
\label{figure:embed noise}
\end{subfigure}

\begin{subfigure}[t]{0.24\textwidth}
\centering
\includegraphics[width=0.52\textwidth]{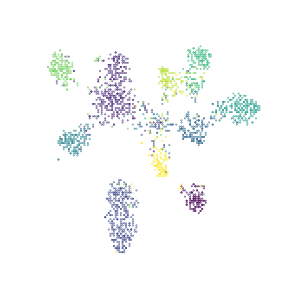}
\caption{t-SNE Integer Rounding}
\label{figure:tsne round int}
\end{subfigure}
\hfill
\begin{subfigure}[t]{0.24\textwidth}
\centering
\includegraphics[width=0.52\textwidth]{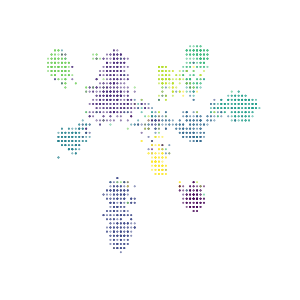}
\caption{t-SNE Even Rounding}
\label{figure:tsne round even}
\end{subfigure}
\hfill
\begin{subfigure}[t]{0.24\textwidth}
\centering
\includegraphics[width=0.52\textwidth]{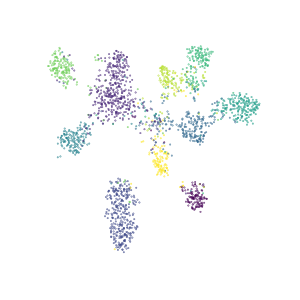}
\caption{t-SNE Noise (2\% STD)}
\label{figure:tsne noise 2}
\end{subfigure}
\hfill
\begin{subfigure}[t]{0.24\textwidth}
\centering
\includegraphics[width=0.52\textwidth]{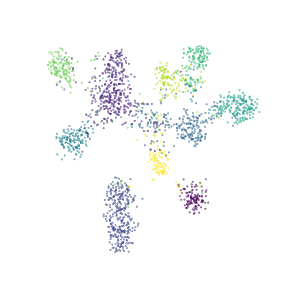}
\caption{t-SNE Noise (5\% STD)}
\label{figure:tsne noise 5}
\end{subfigure}
\caption{t-SNE plots under different defense methods. 
Embedding noise (d) fails visual examination due to much more dispersed clusters. 
t-SNE coordinates rounding (e) and (f) fails due to obvious artifacts (e.g., grid-like patterns).}
\label{figure:tsne defense}
\end{figure*}

\begin{figure*}[hbtp]
\centering
\begin{subfigure}{0.125\textwidth}
\centering
\includegraphics[width=\textwidth]{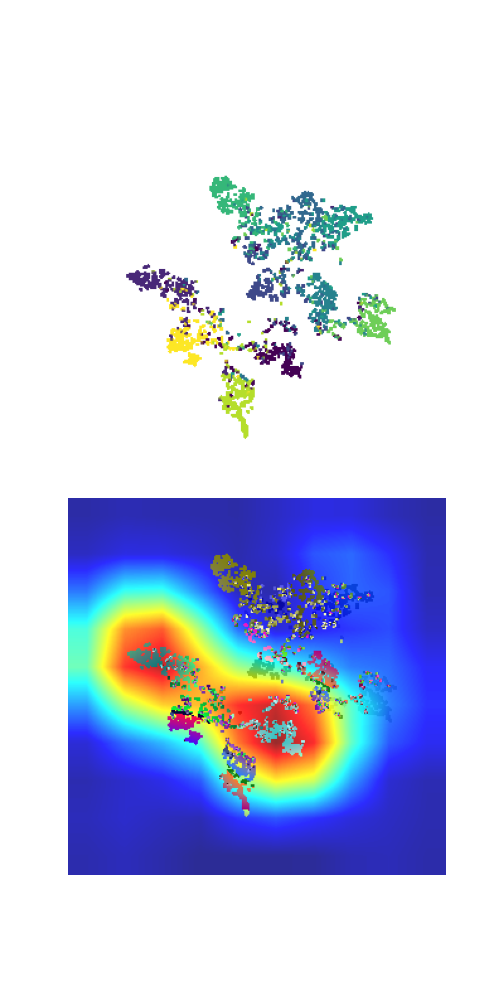}
\caption{ResNet18}
\label{figure:gradcam tsne resnet18}
\end{subfigure}%
\hfill
\begin{subfigure}{0.125\textwidth}
\centering
\includegraphics[width=\textwidth]{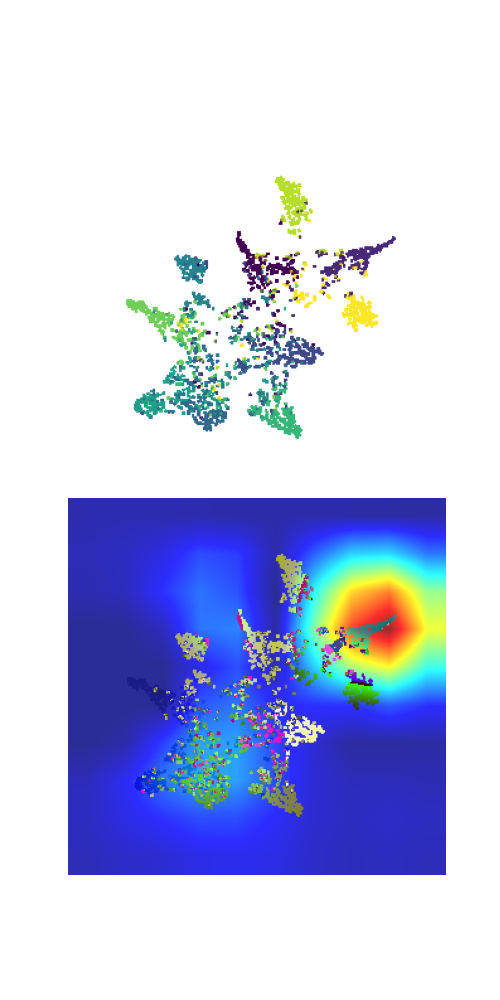}
\caption{ResNet34}
\label{figure:gradcam tsne resnet34}
\end{subfigure}%
\hfill
\begin{subfigure}{0.125\textwidth}
\centering
\includegraphics[width=\textwidth]{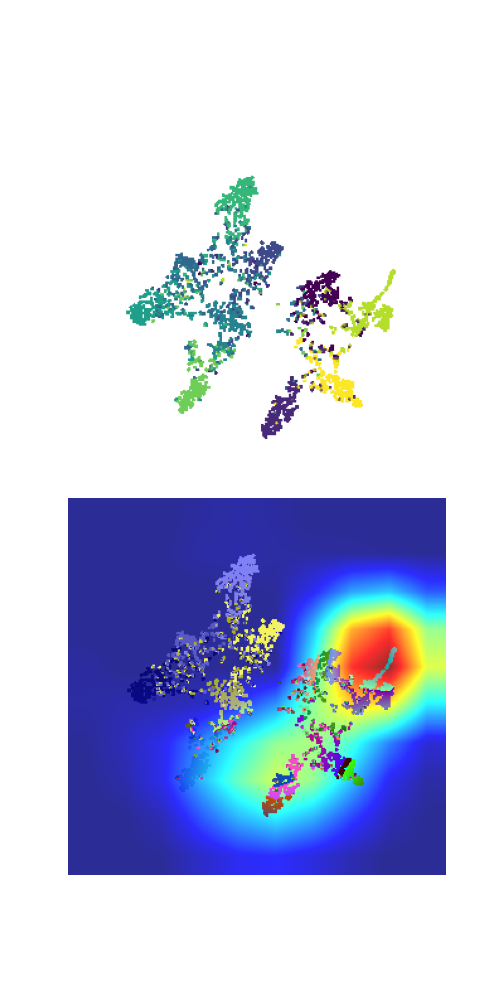}
\caption{ResNet50}
\label{figure:gradcam tsne resnet50}
\end{subfigure}%
\hfill
\begin{subfigure}{0.125\textwidth}
\centering
\includegraphics[width=\textwidth]{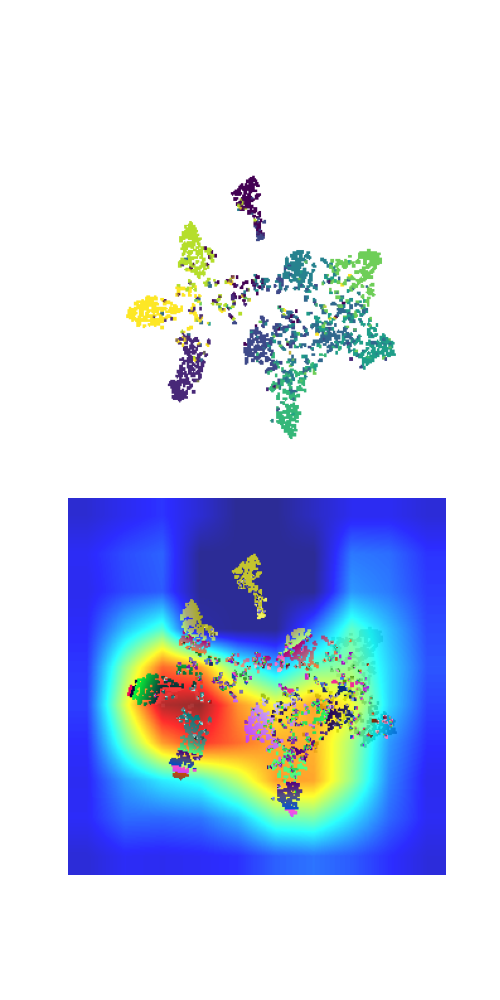}
\caption{MobileV2}
\label{figure:gradcam tsne mobilev2}
\end{subfigure}%
\hfill
\begin{subfigure}{0.125\textwidth}
\centering
\includegraphics[width=\textwidth]{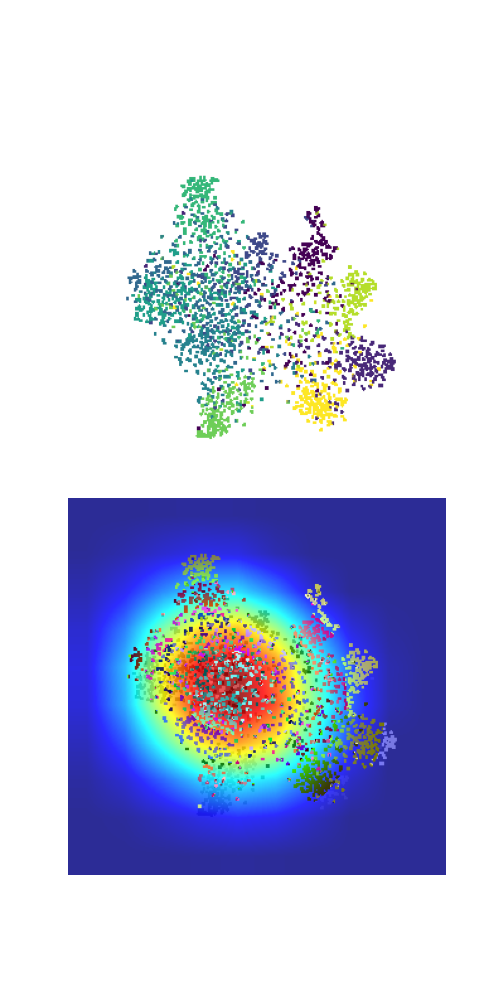}
\caption{MobileV3}
\label{figure:gradcam tsne mobilev3}
\end{subfigure}%
\hfill
\begin{subfigure}{0.125\textwidth}
\centering
\includegraphics[width=\textwidth]{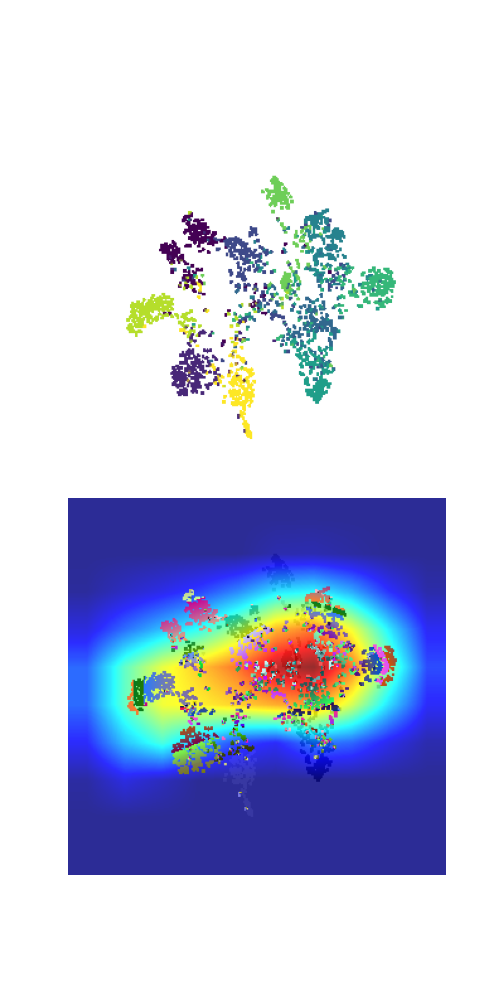}
\caption{DenseNet121}
\label{figure:gradcam tsne densenet}
\end{subfigure}%
\caption{Grad-CAM heat map on t-SNE plots.
The attack model focuses on different patterns.}
\label{figure:gradcam tsne}
\end{figure*}  

\begin{figure*}[hbtp]
\centering
\begin{subfigure}{0.125\textwidth}
\centering
\includegraphics[width=\textwidth]{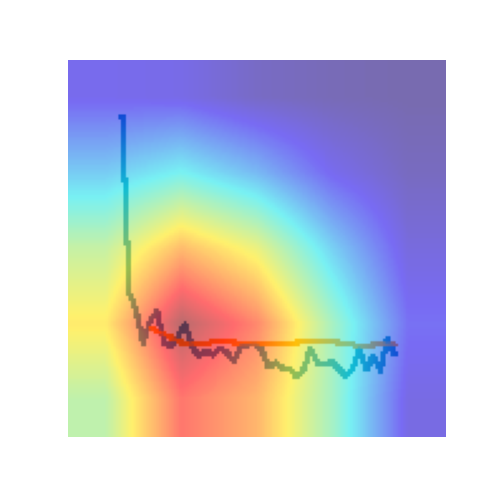}
\caption{ResNet18}
\label{figure:gradcam loss resnet18}
\end{subfigure}%
\hfill
\begin{subfigure}{0.125\textwidth}
\centering
\includegraphics[width=\textwidth]{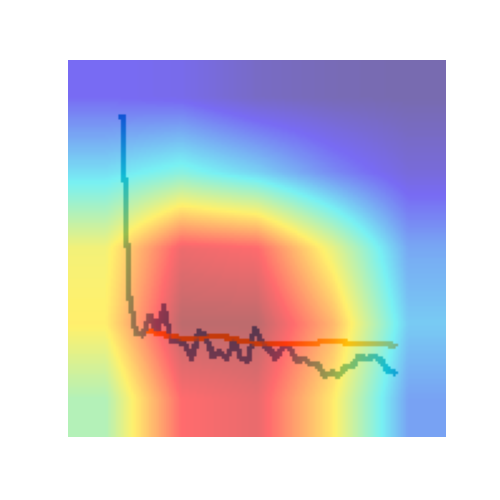}
\caption{ResNet34}
\label{figure:gradcam loss resnet34}
\end{subfigure}%
\hfill
\begin{subfigure}{0.125\textwidth}
\centering
\includegraphics[width=\textwidth]{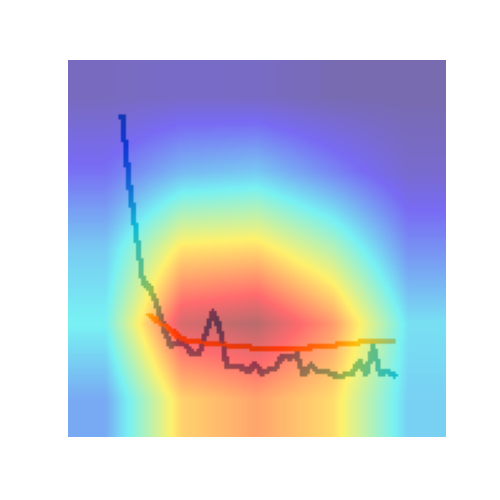}
\caption{ResNet50}
\label{figure:gradcam loss resnet50}
\end{subfigure}%
\hfill
\begin{subfigure}{0.125\textwidth}
\centering
\includegraphics[width=\textwidth]{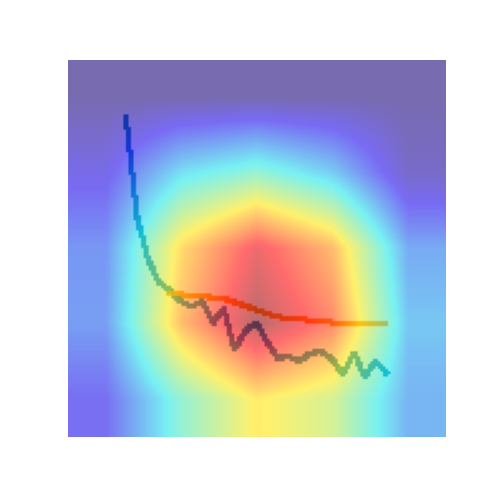}
\caption{MobileV2}
\label{figure:gradcam loss mobilev2}
\end{subfigure}%
\hfill
\begin{subfigure}{0.125\textwidth}
\centering
\includegraphics[width=\textwidth]{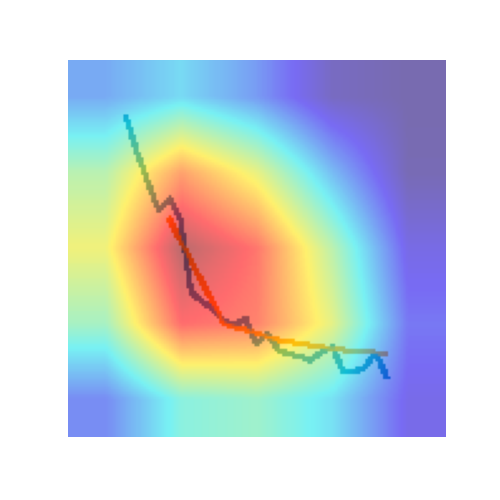}
\caption{MobileV3}
\label{figure:gradcam loss mobilev3}
\end{subfigure}%
\hfill
\begin{subfigure}{0.125\textwidth}
\centering
\includegraphics[width=\textwidth]{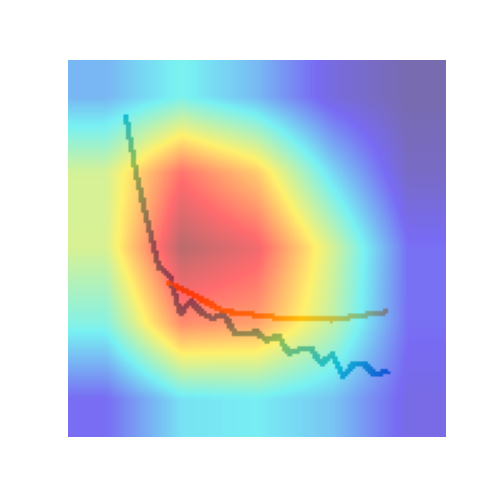}
\caption{DenseNet121}
\label{figure:gradcam loss densenet}
\end{subfigure}%
\caption{The Grad-CAM heat map on loss plots without axis. 
The attack model focuses primarily on early epochs.}
\label{figure:gradcam loss without axis}
\end{figure*}  

\begin{figure*}[hbtp]
\centering
\begin{subfigure}{0.125\textwidth}
\centering
\includegraphics[width=\textwidth]{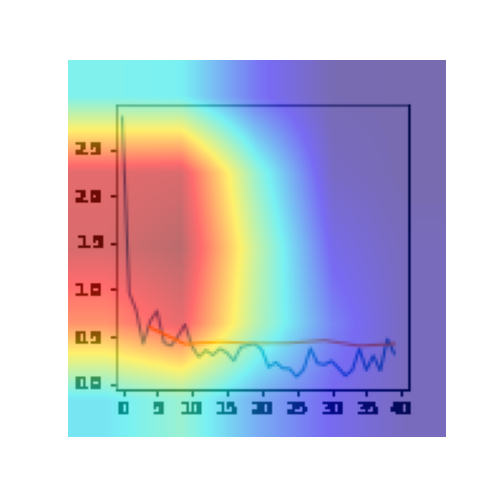}
\caption{ResNet18}
\label{figure:gradcam loss axis resnet18}
\end{subfigure}%
\hfill
\begin{subfigure}{0.125\textwidth}
\centering
\includegraphics[width=\textwidth]{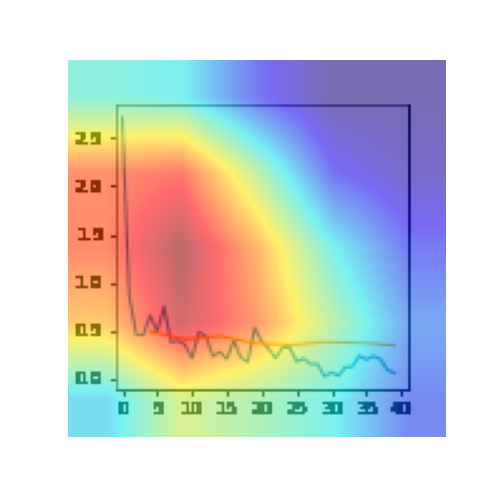}
\caption{ResNet34}
\label{figure:gradcam loss axis resnet34}
\end{subfigure}%
\hfill
\begin{subfigure}{0.125\textwidth}
\centering
\includegraphics[width=\textwidth]{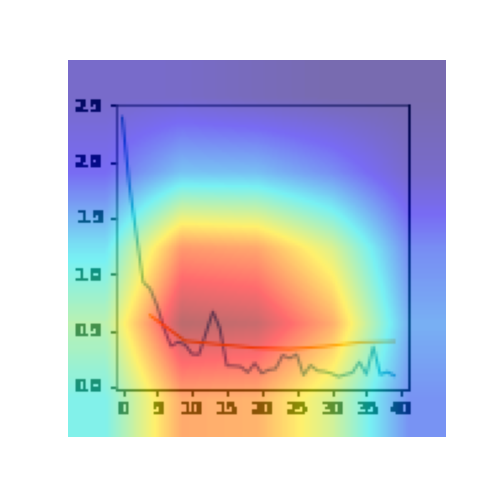}
\caption{ResNet50}
\label{figure:gradcam loss axis resnet50}
\end{subfigure}%
\hfill
\begin{subfigure}{0.125\textwidth}
\centering
\includegraphics[width=\textwidth]{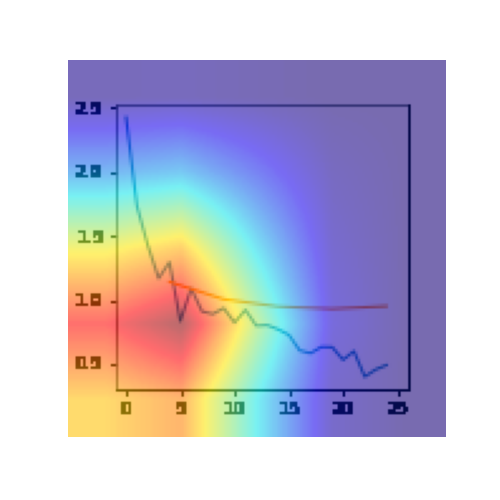}
\caption{MobileV2}
\label{figure:gradcam loss axis mobilev2}
\end{subfigure}%
\hfill
\begin{subfigure}{0.125\textwidth}
\centering
\includegraphics[width=\textwidth]{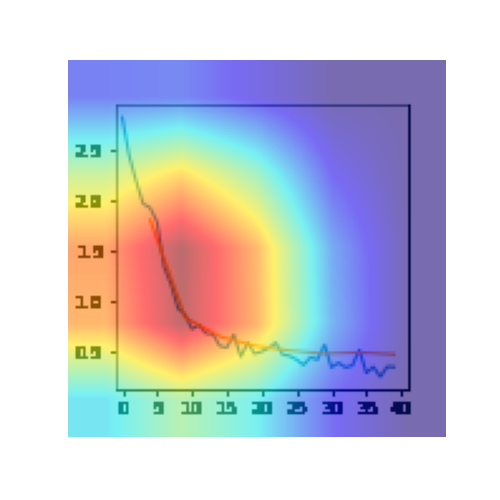}
\caption{MobileV3}
\label{figure:gradcam loss axis mobilev3}
\end{subfigure}%
\hfill
\begin{subfigure}{0.125\textwidth}
\centering
\includegraphics[width=\textwidth]{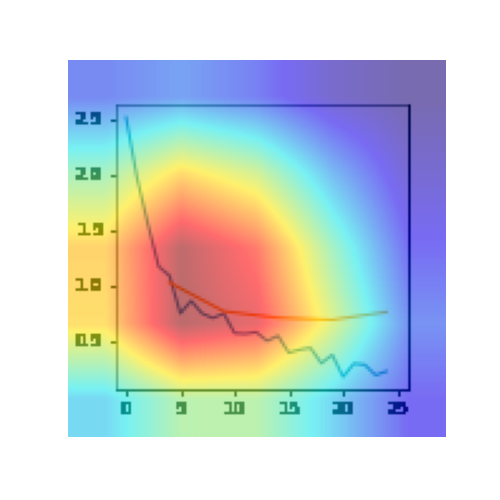}
\caption{DenseNet121}
\label{figure:gradcam loss axis densenet}
\end{subfigure}%
\caption{The Grad-CAM heat map on loss plots with the axis. 
The attack model uses axis information heavily for ResNet18 and ResNet34. 
The added axis information noticeably reduces attack model's misclassification rate within the ResNet family.}
\label{figure:gradcam loss with axis}
\end{figure*}  

\end{document}